\documentclass[prb,article,twocolumn,superscriptaddress]{revtex4-1}
\usepackage{gnuplottex}
\usepackage{mathtools}
\usepackage{float}
\usepackage[english]{babel}
\usepackage[utf8]{inputenc}
\usepackage{amsmath}
\usepackage[table]{xcolor}
\usepackage{amsfonts}
\usepackage{graphicx}
\usepackage[colorlinks=true, linkcolor=blue, citecolor=dred]{hyperref}
\usepackage{algorithm}
\usepackage{algpseudocode}
\usepackage{bbold}
\usepackage{amssymb}
\usepackage{bm}
\usepackage{bbm}
\usepackage{tabularx, booktabs}
\newcolumntype{Y}{>{\centering\arraybackslash}X}

\newcommand\bbone{\ensuremath{\mathbbm{1}}}

\newcommand{\bfn}{\mathbf{n}}

\newcommand{\bfv}{\mathbf{v}}

\definecolor{dgreen}{rgb}{0,.5,0}
\definecolor{dblue}{rgb}{0,0,.5}
\definecolor{dred}{rgb}{0.5,0,.5}

\DeclareMathOperator*{\argmin}{arg\,min}

\begin{document}

\preprint{}

\date{\today}

\title{Projected site-occupation embedding theory}

\author{Bruno Senjean}
\email{bsenjean@gmail.com}
\affiliation{Laboratoire de Chimie Quantique,
Institut de Chimie, CNRS/Universit\'{e} de Strasbourg,
4 rue Blaise Pascal, 67000 Strasbourg, France}
\affiliation{Instituut-Lorentz, Universiteit Leiden, P.O. Box 9506, 2300 RA Leiden, The Netherlands}

\begin{abstract}
Site-occupation embedding theory (SOET) 
[B. Senjean {\it et al.}, Phys. Rev. B {\bf 97}, 235105 (2018)] 
is an in-principle exact embedding method
combining wavefunction theory and density functional theory
that gave promising
results when applied to the one-dimensional
Hubbard model.
Despite its overall good performance, SOET 
faces a computational cost
problem as its
auxiliary impurity-interacting system
remains the size of the full system (which is problematic
as the computational cost increases exponentially with
system size).
In this work,
this issue is circumvented by employing
the Schmidt decomposition, thus leading to
a drastic reduction of the 
computational cost
while retaining the same accuracy.
We show that this projected version of SOET (P-SOET)
is competitive with other
embedding techniques such as density matrix embedding 
theory (DMET)
[G. Knizia and G. K-L. Chan, Phys. Rev. Lett. {\bf 109}, 186404 (2012)]. 
In contrast to the latter, density functional contributions
come naturally into play in P-SOET's framework without any additional computational cost or
double counting effect.
As an important
result, the density-driven Mott--Hubbard transition in one dimension
(which is displayed by multiple impurity sites in DMET or in
dynamical mean-field theory) is well described, for the first time,
with a single impurity site.
\end{abstract}


\maketitle

\section{Introduction}

Due to their unusual and remarkable properties,
strongly correlated materials
play an important role
in the development of innovative technologies,
such as photovoltaic cells, 
pharmaceuticals and industrial catalysts.
These developments
would greatly benefit
from an accurate theoretical description.
Unfortunately, 
the infamous exponential wall problem
prevents the use of correlated wavefunction methods,
and alternatives must be considered.
A major milestone has been delivered 
by Kohn-Sham density functional
theory (KS-DFT)~\cite{hktheo,KS}. 
Despite its in-principle exact
mathematical foundation,
strongly correlated systems still remain challenging for the present
density functional approximations~\cite{cohen2011challenges,
swart2013spin,swart2016spinning}.
Alternatively,
quantum embedding 
theories~\cite{sun2016quantum} have been proposed
and are the main focus of this paper.
The strategy of embedding techniques consists in
solving only a small part of the system
(referred to as the fragment)
by a high-level
method,
while a low-level
approximation is used for the rest of the 
system (referred to as the environment).
Green-function-based methods
have been developed,
such as the widely used
dynamical mean-field theory
(DMFT)~\cite{georges1992hubbard,georges1996limitdimension,
kotliar2004strongly,kotliar2006reviewDMFT,held2007electronic,
vollhardt2012dynamical,martin2016interacting}
or the more recent self-energy embedding
theory (SEET)~\cite{kananenka2015systematically,
lan2015communication,lan2016rigorous,zgid2017finite,
lan2017testing,tran2018spin,rusakov2018self}.
If one is interested about ground-state 
properties only, the Green function can be replaced
by frequency-independent variables, such as the one-particle
reduced density matrix (1RDM) or the electron density.
This has led to the
density matrix embedding theory
(DMET)~\cite{knizia2012density}, the
density embedding theory (DET)~\cite{bulik2014density},
and related methods that
allow
overlapping between 
fragments~\cite{welborn2016bootstrap,ricke2017performance,
ye2018incremental,mordovina2019self}.
They rely on the Schmidt decomposition of the 
full system wavefunction. It generates
an embedded problem sufficiently small to be
solved by
exact
diagonalization~\cite{knizia2012density,knizia2013density,
bulik2014density}, 
density matrix renormalization
group~\cite{wouters2016practical},
auxiliary-field quantum Monte Carlo~\cite{zheng2017cluster}
and more recently the complete active space self-consistent 
field~\cite{pham2018can,hermes2019multiconfigurational}.
In practice, the exact
wavefunction of the full system is not known
and has been
approximated by
Hartree--Fock (HF)~\cite{knizia2012density,knizia2013density},
spin-unrestricted HF~\cite{bulik2014density},
HF-Bogoliubov~\cite{leblanc2015solutions,zheng2016ground},
antisymmetrized geminal power~\cite{tsuchimochi2015density},
block-product states for spin
lattices~\cite{fan2015cluster,
gunst2017block},
multiconfigurational self-consistent 
field~\cite{hermes2019multiconfigurational}
and KS-DFT~\cite{mordovina2019self}.
Note that 
DMET has been mostly applied to model
Hamiltonians~\cite{knizia2012density,
bulik2014density,
leblanc2015solutions,tsuchimochi2015density,
zheng2016ground,
zheng2017cluster,zheng2017stripe,
fan2015cluster,gunst2017block,
sandhoefer2016density,reinhard2019density,
mukherjee2017simple}
but also to quantum
chemical systems~\cite{knizia2013density,wouters2016practical,
pham2018can,hermes2019multiconfigurational}.
Extension to
excited-state 
properties~\cite{chen2014intermediate,booth2015spectral}
and to non-equilibrium electron
dynamics~\cite{kretchmer2018real} have been investigated,
as well as rigorous combinations of DMET with
DMFT~\cite{fertitta2018rigorous}
and with the
rotational invariant slave bosons
theory~\cite{lee2019rotationally}.

In this work, a new method stemming from
site-occupation embedding theory (SOET)~\cite{fromager2015exact,
senjean2017local,senjean2018site,senjean2018multiple,
senjean2018thesis} is
proposed and relies on the Schmidt decomposition. 
In SOET, only the fragment
is explicitly interacting,
in the line of DMET (in its non-interacting bath formulation) and 
other embedding approaches.
But instead of being approximated,
the environment
is described in-principle exactly
by a complementary
functional of the density.
Still in its early
stages, SOET has been applied to the one-dimensional Hubbard 
model although an extension to quantum chemistry is not 
excluded~\cite{fromager2015exact,senjean2018thesis}.
The attractiveness
of SOET resides in its
in-principle exact embedding framework that combines
both wavefunction theory (WFT) 
and DFT without double counting problem. However,
it has remained prohibitively
expensive, as the size of the auxiliary impurity-interacting 
system remained the size of the original (fully-interacting) problem.
In this work, we take advantage of
the Schmidt decomposition to rigorously
remove most of the degrees of freedom in the environment, and
to replace it by a bath that is the same size as the fragment 
(i.e. the impurity-interacting sites).
We refer to this new method as projected-SOET (P-SOET), and apply
it to the one-dimensional uniform Hubbard model up to 400 sites.
We show that P-SOET yields
accurate results
at a drastically reduced computational cost.

The paper is organized as follows. 
After a brief summary of the key equations of SOET in 
Sec.~\ref{sec:SOET}, P-SOET and its implementation
are introduced in Sec.~\ref{sec:PSOET}.
Details on the functionals of the density are
given in Sec.~\ref{sec:approx} followed by the computational details in 
Sec.~\ref{sec:comp_details}.
Results obtained on the one-dimensional uniform Hubbard model
are provided in Sec.~\ref{sec:results}. The
density-driven Mott-transition is studied in Sec.~\ref{sec:transition}. Finally,
conclusions and perspectives are given in Sec.~\ref{sec:conclu}.

\section{Theory}

\subsection{Site-occupation embedding theory}\label{sec:SOET}

For the paper to be self-contained, this section summarizes the key 
equations of SOET~\cite{fromager2015exact,
senjean2017local,senjean2018site,senjean2018multiple}.
We start with the (not necessarily uniform) one-dimensional $L$-site 
Hubbard model in an external potential ($\mathbf{v} 
\equiv \lbrace v_i \rbrace_i$),
\begin{eqnarray}\label{eq:hubbard_H}
\hat{H} = -t \sum_{i=0,\sigma}^{L-1} \left(
\hat{c}_{i\sigma}^\dagger \hat{c}_{i+1\sigma} + h.c.\right)
+ U \sum_{i=0}^{L-1} \hat{n}_{i\uparrow} \hat{n}_{i\downarrow}
+ \sum_{i=0}^{L-1} v_i \hat{n}_{i}, \nonumber \\
\end{eqnarray}
with $t>0$ the hopping parameter between first neighbor sites and
$U$ the on-site electronic repulsion. The
site-occupation operator
$\hat{n}_i = \hat{n}_{i\uparrow} + \hat{n}_{i\downarrow}$
(with $\hat{n}_{i\sigma} = 
\hat{c}_{i\sigma}^\dagger \hat{c}_{i\sigma}$) is expressed using the
creation ($\hat{c}_{i\sigma}^\dagger$) and annihilation ($\hat{c}_{i
\sigma}$) operators of an electron of spin $\sigma$ on site $i$.
The ground-state energy of this model can be expressed in 
site-occupation functional theory 
(SOFT)~\cite{gunnarsson1986density,DFT_lattice}
(DFT analog for model
Hamiltonians)
using the following
variational principle:
\begin{eqnarray}\label{eq:ener_min_n}
E(\mathbf{v})= \underset{\mathbf{n}}{\rm min} 
\left \lbrace F(\mathbf{n}) + 
(\mathbf{v}| \mathbf{n}) \right \rbrace,
\end{eqnarray}
where
\begin{eqnarray}
F(\mathbf{n}) = \underset{\Psi 
\rightarrow \mathbf{n}}{\rm min} \left 
\lbrace \left\langle \Psi \middle\vert \hat{T} +
\hat{U}
\middle\vert
\Psi \right\rangle \right 
\rbrace
 \label{eq:LL}
\end{eqnarray}
is the Levy-Lieb (LL) functional
and $(\mathbf{v}| \mathbf{n}) = \sum_i v_i n_i$.
The idea of SOET relies on mapping the fully interacting problem
onto a partially interacting one. The interacting sites are referred to as
{\it impurities}, and $M$ impurities are embedded in a
{\it bath}
of $(L-M)$ noninteracting sites.
(Note that the term {\it bath} in SOET actually refers to the
{\it environment} in DMET.)
This is done by decomposing the conventional LL
functional
as follows~\cite{senjean2018multiple}:
\begin{eqnarray}
F(\bfn) = F_M^{\rm imp}(\bfn) + 
\overline{E}_{{\rm Hxc},M}^{\rm bath}(\bfn),
\end{eqnarray}
where
\begin{eqnarray}
F_M^{{\rm imp}}(\mathbf{n}) = \underset{\Psi 
\rightarrow \mathbf{n}}{\rm min} \left 
\lbrace \left\langle \Psi \middle\vert \hat{T} +
\hat{U}_M
\middle\vert
\Psi \right\rangle \right 
\rbrace
 \label{eq:LL_Mimp}
\end{eqnarray}
is the $M$-impurity-interacting functional with
$\hat{U}_M=U\sum^{M-1}_{i=0}
\hat{n}_{i\uparrow}\hat{n}_{i\downarrow}$,
and
$\overline{E}_{{\rm Hxc},M}^{\rm bath}(\bfn)$ is the complementary
Hxc bath energy, functional of the
sites occupation.
Inserting this decomposition into 
Eq.~(\ref{eq:ener_min_n}) leads to the
variational principle within SOET~\cite{senjean2018multiple}:
\begin{eqnarray}
\label{eq:variational_energy}
E(\mathbf{v})= 
\underset{\Psi}{\rm min} \Big 
\lbrace
&& \left\langle \Psi \middle\vert \hat{T} +
\hat{U}_M
\middle\vert
\Psi \right\rangle 
+\overline{E}_{{\rm Hxc},M}^{\rm 
bath}\left(\mathbf{n}^\Psi\right)
+ 
\left(\mathbf{v}| \mathbf{n}^\Psi\right)
\Big\rbrace
,
\nonumber\\
\end{eqnarray}
where
$\mathbf{n}^\Psi\equiv\left\{\langle\Psi\vert\hat{n}_i
\vert\Psi\rangle\right\}_i$.
The
minimizing
$M$-impurity-interacting wavefunction
$\Psi_M^{\rm imp}$ in Eq.~(\ref{eq:variational_energy}) reproduces the
exact density
profile of the fully-interacting system
and fulfills the following 
self-consistent equation:
\begin{eqnarray}\label{eq:self-consistent-SOET_new}
&& \left( -t \sum_{i=0,\sigma}^{L-1} \left( \hat{c}_{i\sigma}^\dagger \hat{c}_{i
+1\sigma} + {\rm H.c.}\right) + \hat{U}_{M} +
 \displaystyle \sum_{i=0}^{L-1}  v_{M,i}^{\rm emb} \hat{n}_i \right)\vert \Psi_M^{\rm imp} \rangle  \nonumber \\
&& = \mathcal{E}_M^{\rm imp} \vert \Psi_M^{\rm imp} \rangle, 
\end{eqnarray}
where $\mathcal{E}^{\rm imp}_M$
is the impurity auxiliary energy (i.e., the ground-state energy of the $M$-impurity interacting Hamiltonian) and
\begin{eqnarray}\label{eq:embedding_pot}
v^{\rm emb}_{M,i} =v_i+\dfrac{\partial  \overline{E}^{\rm bath}_{{\rm
Hxc},M}(\mathbf{n}^{\Psi^{\rm imp}_M})}{\partial n_i}
\end{eqnarray}
corresponds to the embedding
potential for the $M$ impurities embedded into the $(L-M)$ bath sites.
Concerning the complementary Hxc bath functional,
it can be shown that~\cite{fromager2015exact}
\begin{eqnarray}
\label{eq:Ecbath_expression}
\overline{E}_{{\rm c},M}^{\rm bath}(\mathbf{n})=E_{\rm c}(\mathbf{n}) -
E_{{\rm 
c},M}^{\rm imp} (\mathbf{n}),
\end{eqnarray}
where $E^{\rm imp}_{{\rm c},M}(\bfn)$ is the analog of the correlation
functional for the $M$-impurity-interacting
system.
Turning to the uniform problem investigated 
in this paper [$\bfv = 0$ in Eq.~(\ref{eq:hubbard_H})],
the local density approximation (LDA)
of $E_{\rm c}(\bfn)$ is exact and reads
\begin{eqnarray}
E_{\rm c}(\bfn) = \sum_i e_{\rm c}(n_i),
\end{eqnarray}
where $e_{\rm c}(n)$ is the per-site correlation energy functional.
In addition, the exact per-site
energy $e = E(\mathbf{v}=\mathbf{0})/L$ and double occupation expressions for the uniform Hubbard 
model have been derived in 
Ref.~\cite{senjean2018multiple} and read
\begin{eqnarray}\label{eq:per_site_ener}
e&=&
\dfrac{1}{M}\left.
\sum^{M-1}_{i=0}
\left[
t_{\rm s}(n_i^\Psi)+t\dfrac{\partial e_{\rm c}(n_i^\Psi)}{\partial
t}
+{U}
\langle\hat{n}_{i\uparrow}\hat{n}_{i\downarrow}\rangle_{\Psi}
\right]
\right|_{\Psi=\Psi_M^{\rm imp}}
\nonumber\\
&&
+\left.U
\dfrac{\partial \overline{e}_{{\rm c},M}^{\rm
bath}({\bf n}^\Psi)}{\partial U}
\right|_{\Psi={\Psi_M^{\rm imp}}}
,
\end{eqnarray}
and
\begin{eqnarray}\label{eq:dblocc_SOET}
d=\dfrac{1}{M}
\sum^{M-1}_{i=0}
\langle\hat{n}_{i\uparrow}\hat{n}_{i\downarrow}\rangle_{\Psi_M^{\rm imp}}
+\dfrac{\partial \overline{e}_{{\rm c},M}^{\rm
bath}({\bf n}^{\Psi_M^{\rm imp}})}{\partial U},
\end{eqnarray}
respectively,
where $t_s(n)=-4t\sin(\pi n/2)/\pi$ is the (one-dimensional) per-site non-interacting kinetic energy
functional and
\begin{eqnarray}\label{eq:ecbath_per_site_M_uniform}
\overline{e}_{{\rm c},M}^{\rm
bath}(\mathbf{n})=\dfrac{1}{M}\left[\left(\sum^{M-1}_{i=0}e_c(n_i)\right)-E_{{\rm 
c},M}^{\rm imp} (\mathbf{n})\right]
\end{eqnarray}
is the per-site analog of the bath correlation energy in 
Eq.~(\ref{eq:Ecbath_expression}).
Interestingly, the per-site energy and the double occupation are
related to each other as follows:
\begin{eqnarray}\label{eq:per_site_dblocc}
e&=&
U d + 
\dfrac{1}{M}\left.
\sum^{M-1}_{i=0}
\left[
t_{\rm s}(n_i^\Psi)+t\dfrac{\partial e_{\rm c}(n_i^\Psi)}{\partial
t} \right]\right|_{\Psi={\Psi_M^{\rm imp}}}.
\end{eqnarray}

\subsection{Projected site-occupation embedding theory}\label{sec:PSOET}

In SOET, the auxiliary impurity-interacting
problem has the same size as the
full system even though the interactions are restricted
to the fragment of interest.
This is the actual bottleneck of
SOET.
Indeed, as currently implemented,
considering the kinetic operator in the whole system
in Eq.~(\ref{eq:self-consistent-SOET_new}) does not lead to any 
reduction of the computational cost compared to solving
the fully-interacting system.
In the following, we describe how
the Schmidt decomposition is employed in P-SOET 
to drastically reduce the size 
of the system to which a correlated wavefunction theory is applied.
First, we review the Schmidt decomposition
introduced by Knizia and Chan within the now
well-established
DMET~\cite{knizia2012density}. We then discuss
how to derive the embedded
problem in P-SOET.

\subsubsection{Review of the Schmidt decomposition}\label{sec:schmidt}

Suppose that 
the system is divided into a fragment $F$ and the rest of the 
system (referred to as the ``environment'') $E$.
For instance, regarding the SOET Hamiltonian in 
Eq.~(\ref{eq:self-consistent-SOET_new}),
the fragment is composed
of the $M$ explicitly interacting impurity sites while
the remaining
implicitly interacting sites correspond to the {\it environment}.
Note that in SOET, the latter is referred to as the {\it bath}, but
in this paper we will follow the DMET nomenclature where the {\it bath}
states are obtained after the Schmidt decomposition. 
The total Hilbert
space
of the system is the tensor product of 
$F$ and $E$,
$\mathcal{H} = \mathcal{H}_F \otimes \mathcal{H}_E$,
where $\mathcal{H}_F$ ($\mathcal{H}_E$)
has size $N_F = d^{L_F}$ ($N_E = d^{L_E}$).
$L_F$ ($L_E$) is the number of sites in subsystem $F$ ($E$).
Given that a site is equivalent to a spatial orbital,
there are $d=4$ possible occupations:
empty, singly occupied with spin-projection $s^z = \pm 1/2$, 
and doubly occupied
$\lbrace \vert 0 \rangle, \vert \uparrow \rangle, 
\vert \downarrow \rangle, 
\vert \uparrow\downarrow \rangle \rbrace$.
Let $\lbrace \vert F_i \rangle\rbrace_i$ 
($\lbrace \vert E_j \rangle\rbrace_j$)
denote the many-body basis of size 
$N_F$ ($N_E$).
$\mathcal{H}$ is then spanned by 
$N_F N_E = 4^{L_F + L_E} = 4^L$ 
many-body states denoted by
$\lbrace \vert F_i \rangle \vert E_j \rangle \rbrace_{ij}$. 
The ground state in $\mathcal{H}$
can be expressed as
\begin{eqnarray}
\vert \Psi_0 \rangle = \sum_{i=1}^{N_F} \sum_{j=1}^{N_E} C_{ij} 
\vert F_i \rangle 
\vert E_j \rangle.
\end{eqnarray}
Performing the singular value 
decomposition and
assuming $N_F < N_E$ leads to
\begin{eqnarray}
\vert \Psi_0 \rangle & = & \sum_{i=1}^{N_F} \sum_{j=1}^{N_E} 
\sum_{n=1}^{N_F} 
U_{in} \lambda_n V^\dagger_{n j} \vert F_i \rangle \vert E_j 
\rangle 
\nonumber \\
& = & \sum_{n=1}^{N_F} 
\lambda_n \vert \tilde{F}_n \rangle \vert \tilde{B}_n 
\rangle ,
\label{Schmidt}
\end{eqnarray}
where $U_{in}$ and $V_{jn}^* = V_{n j}^\dagger$ 
rotate the many-body basis 
$\lbrace \vert F_i \rangle \rbrace_i$ and
$\lbrace \vert E_j \rangle \rbrace_j$ into 
a new many-body basis
$\lbrace \vert \tilde{F}_n\rangle \rbrace_n$ and 
$\lbrace \vert \tilde{B}_n\rangle 
\rbrace_n$ (where $\tilde{B}$ now refers to the new {\it bath} states, 
to be distinguished with the environment states).
Eq.~(\ref{Schmidt}) is called the 
Schmidt decomposition of the wavefunction~\cite{knizia2012density},
now expressed within the
reduced Hilbert space spanned by
$N_F^2 = 4^{2L_F}$
many-body states 
(denoted by
$\lbrace \vert \tilde{F}_m \rangle\vert \tilde{B}_n \rangle \rbrace_{mn}$), 
thus removing most of the degrees of freedom in the environment.
The corresponding embedded Hamiltonian is then
obtained by {\it projection},
\begin{eqnarray}
\hat{H}^{\rm emb} = P^\dagger \hat{H} P,
\end{eqnarray}
where $P$ defines the projector onto the Schmidt basis,
\begin{eqnarray}
P = \sum_{m=1,n=1}^{N_F} \vert \tilde{F}_m \rangle \vert \tilde{B}_n  \rangle \langle \tilde{B}_n \vert \langle \tilde{F}_m \vert.
\end{eqnarray}
This projector does not affect the exact ground-state wavefunction,
$P\vert \Psi_0 \rangle = \vert \Psi_0 \rangle$,
such that
\begin{eqnarray}
\hat{H} \vert \Psi_0 \rangle = E_0 \vert \Psi_0 \rangle \rightarrow
\hat{H}^{\rm emb} \vert \Psi_0 \rangle = E_0 \vert \Psi_0 \rangle.
\end{eqnarray}
The ground state of the embedded Hamiltonian is then also
the ground state of the full system.
However, this exact decomposition requires the 
\textit{a priori} knowledge of the exact 
wavefunction, which is of course not known.
An approximate wavefunction has to be used instead
and the single Slater determinant obtained from
a KS-SOFT calculation is considered in this work,
thus leading to approximate single-particle bath states
$\vert \tilde{B}_n \rangle$.

\subsubsection{P-SOET}\label{sec:lattice_embedded}

Let us consider the fully-interacting lattice problem given
by the $L$-site uniform Hubbard 
Hamiltonian.
The first step is to obtain the approximate single-particle
bath states defining the
projection operator from a non-interacting system as reference.
In DMET, this reference commonly corresponds to the 
mean-field approximation of the physical system.
In P-SOET, it makes sense to consider a non-interacting system
which has the same density as the physical one,
as this density will
be used in the determination of the per-site energy
and the double occupation [Eqs.~(\ref{eq:per_site_ener}) and 
(\ref{eq:dblocc_SOET})]. Hence, we consider the
KS-SOFT
Hamiltonian,
\begin{eqnarray}\label{eq:h_soft}
\hat{h}^{\rm KS} = -t \sum_{i=0,\sigma}^{L-1} \left( \hat{c}_{i\sigma}^\dagger \hat{c}_{i
+1\sigma} + {\rm H.c.}\right) + \sum_{i=0}^{L-1} \dfrac{\partial E_{\rm Hxc}(\bfn)}{\partial n_i}\hat{n}_i, \nonumber \\
\end{eqnarray}
because its (self-consistently determined)
effective potential reproduces
the
same density as the physical system,
in principle exactly. Note that this choice has
also been made 
recently by Mordovina {\it et al.}~\cite{mordovina2019self}.
The obtained
density is then inserted
into Eq.~(\ref{eq:embedding_pot}) to determine the
analytical
embedding potential, used to construct
the one-body effective Hamiltonian
\begin{eqnarray}\label{eq:h_eff}
\hat{h}^{\rm eff}= -t \sum_{i=0,\sigma}^{L-1} \left(
\hat{c}_{i\sigma}^\dagger \hat{c}_{i+1\sigma} + {\rm H.c.}\right) 
+ \sum_{i=0}^{L-1} v_{M,i}^{\rm emb}\hat{n}_i, \nonumber \\
\end{eqnarray}
which is nothing but the one-body part of the SOET 
Hamiltonian in Eq.~(\ref{eq:self-consistent-SOET_new})
($\hat{H}^{\rm SOET} = \hat{h}^{\rm eff} + \hat{U}_M$).

The one-body part of the embedded Hamiltonian
is then obtained by projecting 
the one-body effective Hamiltonian
[Eq.~(\ref{eq:h_eff})] as follows:
\begin{eqnarray}\label{eq:h_emb}
\hat{h}^{\rm emb} = P^\dagger \hat{h}^{\rm eff}P,
\end{eqnarray}
where the projector $P$ [of size $(L \times 2L_F)$]
comes from the
Schmidt decomposition of the ground state $\Phi_{\rm KS}$ of 
$\hat{h}^{\rm KS}$~\footnote{We have tried to use a different 
projector generated from the ground-state of the effective Hamiltonian
[Eq.~(\ref{eq:h_eff})] instead of the KS Hamiltonian 
[Eq.~(\ref{eq:h_soft})], but it led to insufficiently accurate results.} 
and reads~\cite{ayral2017dynamical}
\begin{eqnarray}\label{eq:DMET_projector}
P = \begin{bmatrix}
\bbone & 0 \\
0 & C_BC_F^\dagger
\end{bmatrix}
.
\end{eqnarray}
$\bbone$ denotes the identity matrix of size
$(L_F \times L_F)$ and
$C_BC_F^\dagger$ is
a $(L_E \times L_F)$ rectangular matrix which 
is the transformation from the environment to the bath. The reader is
referred to Ref.~\cite{ayral2017dynamical} for a detailed 
construction of this projector.
As readily seen in Eq.~(\ref{eq:DMET_projector}), the transformation
of the one-particle fragment states in the
original basis is the identity.
The fragment is therefore invariant under this projection.
The on-site electron repulsion is then added to the
fragment (or impurity) sites thus leading to the following many-body
embedded Hamiltonian:
\begin{eqnarray}\label{eq:H_emb}
\hat{H}^{\rm emb} = \sum_{ij}^{L_F + L_B} h^{\rm emb}_{ij} 
\left( \hat{c}_{i\sigma}^\dagger \hat{c}_{j\sigma} + 
{\rm H.c.}\right) + U \sum_{i}^{L_F} \hat{n}_{i\uparrow} 
\hat{n}_{i\downarrow}, \nonumber \\
\end{eqnarray}
where $L_F$ (equivalent to $M$ introduced previously) and
$L_B$ are the number of
fragment sites and bath sites, respectively, and $L_F = L_B$.
We refer to $\Psi^{\rm emb}_M$ as its
associated ground-state wavefunction.
The {\it closed} embedded subsystem (fragment+bath) is then twice 
the size of the fragment 
and consists of $4L_F$ spin-orbitals (or $2L_F$ sites)
with $2L_F$ electrons.
One can therefore see the fragment as being an {\it open} system
with the bath playing the role of a reservoir, i.e. the fragment
can contain a fractional number of electrons~\cite{knizia2013density}.
This is a drastic simplification of SOET as well as 
an alternative to DMET
where the correlation potential is now functional of the density.
From $\Psi^{\rm emb}_M$,
we extract the impurity double occupations
\begin{eqnarray}\label{eq:dimp_emb}
\langle\hat{n}_{i\uparrow}\hat{n}_{i\downarrow}\rangle_{\Psi^{\rm emb}_M}, ~~ 0 \leqslant i \leqslant M-1.
\end{eqnarray}
Together with the KS-SOFT density, they are used
to determine the physical per-site energy and double
occupation in Eqs.~(\ref{eq:per_site_ener}) and 
(\ref{eq:dblocc_SOET}), respectively.
Another significant 
difference with DMET
is that no matching condition
(known to be a bottleneck of DMET~\cite{wu2019projected}) 
between the 1RDM of the low-level and the 
high-level wavefunctions is requested in P-SOET.
Instead, only the in-principle exact
density (from the low-level KS determinant) is determined
self-consistently, while the high-level embedded wavefunction
together with
complementary functionals of the density allows for an accurate
evaluation of physical observables.
Note that one can
determine the density of the impurity sites
\begin{eqnarray}\label{eq:occ_emb}
n_i^{\Psi^{\rm emb}_M} = \langle \hat{n}_{i\uparrow} + 
\hat{n}_{i\downarrow} \rangle_{\Psi^{\rm emb}_M}, 
~~ 0 \leqslant i \leqslant M-1
\end{eqnarray}
self-consistently, and use it
instead of the KS-SOFT density.
For the sake of conciseness, this alternative procedure is
described in Appendix~\ref{sec:self-cons}.

\begin{figure*}
\centering
\resizebox{\textwidth}{!}{
\includegraphics[scale=1]{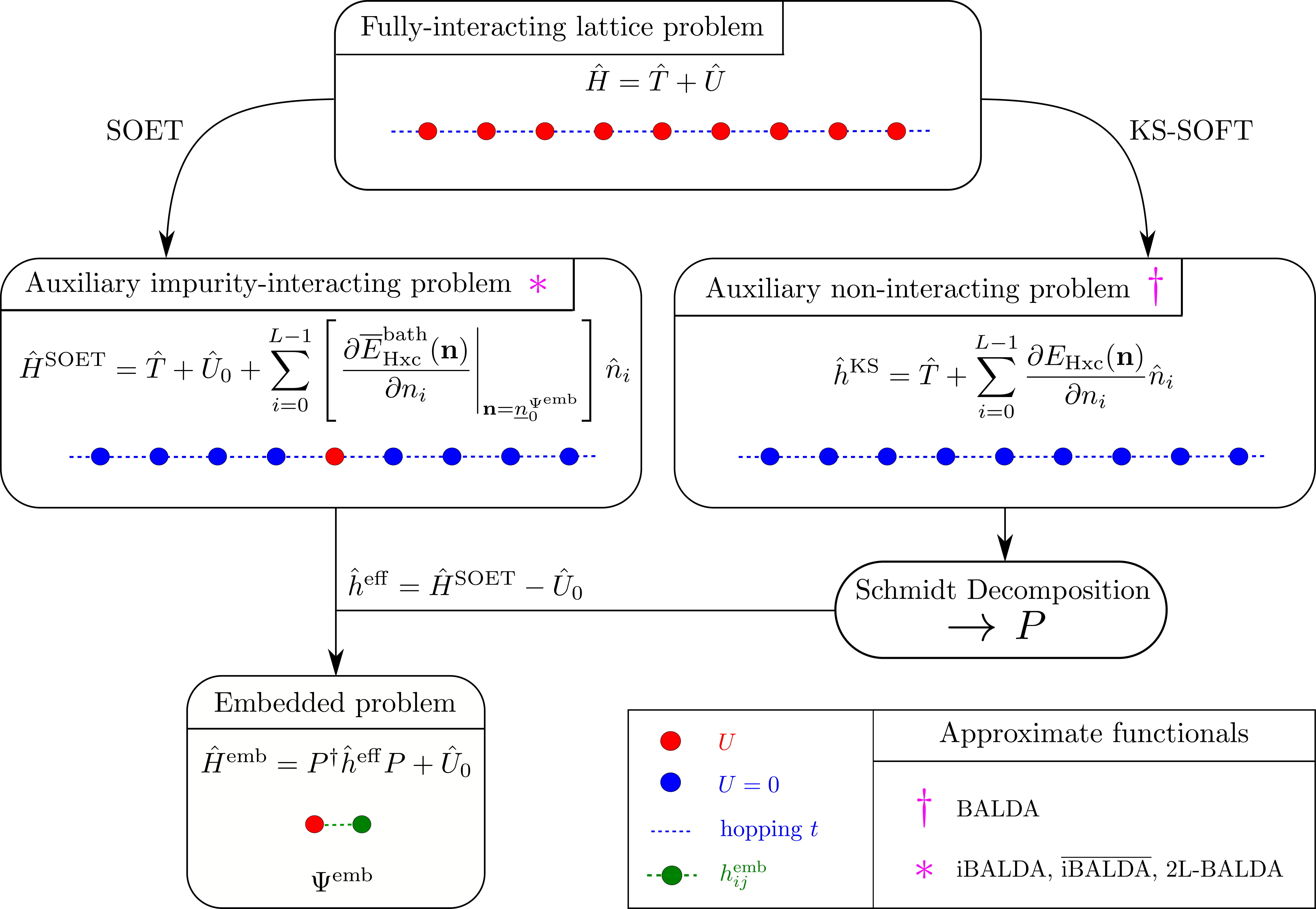}
}
\caption{Representation of the P-SOET algorithm for a 
single impurity site. 
In SOET, the 
auxiliary impurity-interacting problem has to be solved entirely.
In P-SOET, it is projected onto a much smaller embedded 
problem thanks to the Schmidt decomposition of the 
auxiliary non-interacting problem.
An optional self-consistent loop can be added to update the impurity occupation obtained by solving the embedded problem (see Appendix.~\ref{sec:self-cons} and Fig.~\ref{fig:PSOET_selfcons}).
}
\label{fig:PSOET}
\end{figure*}

The procedure detailed in this section
can be summarized as follows:
\begin{enumerate}
\item Solve the KS-SOFT problem [Eq.~(\ref{eq:h_soft})]
\item Apply the Schmidt decomposition to the KS determinant
and determine the 
projection operator $P$
\item Determine the one-body effective Hamiltonian 
[Eq.~(\ref{eq:h_eff})] 
with the density obtained from Step 1
\item Project the effective Hamiltonian to obtain the one-body
embedded Hamiltonian [Eq.~(\ref{eq:h_emb})]
\item Define the embedded Hamiltonian by adding interaction on the 
fragment sites [Eq.~(\ref{eq:H_emb})]
\item Solve the embedded problem to get the embedded wavefunction
$\Psi^{\rm emb}_M$
\end{enumerate}
These steps define the P-SOET algorithm,
pictured in Fig.~\ref{fig:PSOET} for a single
impurity site.

\subsection{Approximate functionals}\label{sec:approx}

The performance of SOET and P-SOET relies
on the accuracy of the per-site correlation
functional $e_{\rm c}(n)$ and
the impurity correlation functional 
$E_{{\rm c},M}^{\rm imp}(\bfn)$,
as well as their derivatives with respect to $U$, $t$ and 
$n_i$~\cite{senjean2018multiple}.
The LDA based on Bethe Ansatz (BA) (so-called BALDA) is 
used for $e_{\rm c}(n)$~\cite{lima2002density,lima2003density,capelle2003density},
and is exact in the thermodynamic limit for
$U = 0$, $U \rightarrow + \infty$ and $n = 1$ for any $U$.
For the impurity correlation functional, a two-level (2L)
approximation based on the Anderson dimer
has been derived in Ref.~\cite{senjean2017local}.
Together with BALDA, it leads to the 
so-called 2L-BALDA 
functional~\cite{senjean2018multiple}, 
 which can be used in the single-impurity case only.
Alternatively, one can use BALDA for the impurity functional as well.
This simple approximation enables us to consider more impurities
and is called the $M$-impurity BALDA 
[iBALDA($M$)]~\cite{senjean2018multiple}. By construction
[see Eq.~(\ref{eq:ecbath_per_site_M_uniform})],
$\overline{e}_{{\rm c},M}^{\rm bath}(\bfn)$ and its
derivatives
are equal to 0
within iBALDA($M$).
The reader is referred to Ref.~\cite{senjean2018multiple} 
for more details on these functionals and their derivations.

\begin{figure}
\centering
\resizebox{\columnwidth}{!}{
\includegraphics[scale=1]{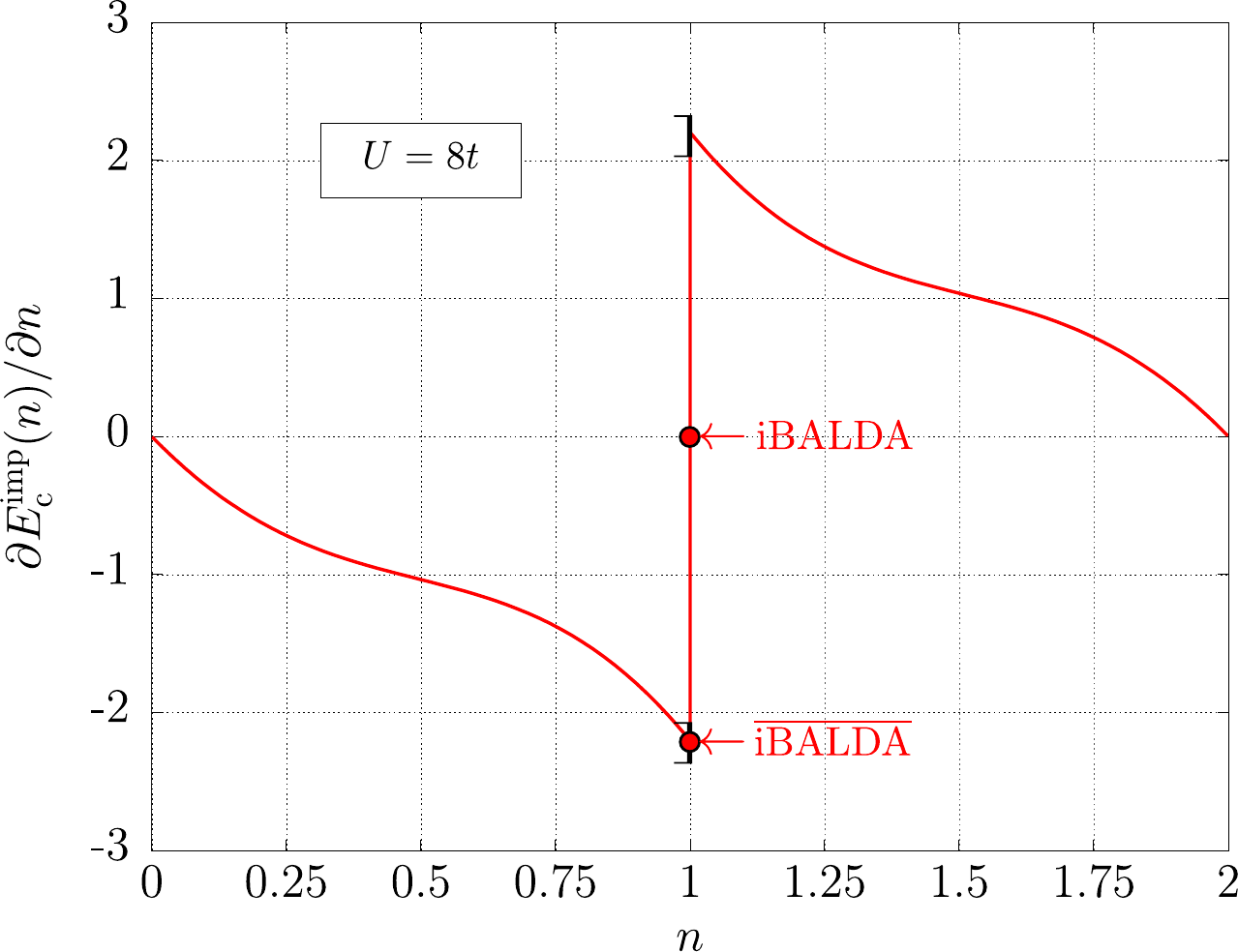}
}
\caption{BALDA correlation 
potential with respect to 
$n$ for $U = 8t$. At half-filling ($n=1$), the iBALDA impurity correlation potential is
exact for a finite system and is equal to 0,
while the $\overline{\rm iBALDA}$ impurity correlation potential is not
and is given by
Eq.~(\ref{eq:BALDA_n=1_pot}).}
\label{fig:iBALDA}
\end{figure}

In SOET,
the exact embedding potential for a finite and half-filled
($\bfn = \underline{1}$)
system,
\begin{eqnarray}\label{eq:emb_pot_HF}
v^{\rm emb}_{M,i}(\bfn = \underline{1}) = -\dfrac{U}{2}
\sum_{j=0}^{M-1} \delta_{ij},  
\end{eqnarray}
has always
been used~\cite{senjean2018site,senjean2018multiple}.
While Eq.~(\ref{eq:emb_pot_HF}) is always fulfilled by
the 2L-BALDA functional,
it is not by iBALDA. Indeed, iBALDA
has been derived from the thermodynamic limit 
($L \rightarrow + \infty$),
for which a derivative discontinuity in the 
correlation energy
functional appears at 
half-filling~\cite{lima2002density,capelle2003density,
senjean2018site} such that the correlation potential is not defined anymore.
Eq.~(\ref{eq:emb_pot_HF}) has then been enforced in previous works.
In this work, another choice is made.
The BALDA (or equivalently, iBALDA) correlation 
potential~\cite{lima2002density,lima2003density,capelle2003density},
\begin{eqnarray}
 \dfrac{\partial e_{\rm c}^{\rm BALDA}(n < 1)}{\partial n} &=&  - 
 2t
 \cos \left( \dfrac{\pi n}{\beta(U/t)}\right) + 
 2t \cos \left( \dfrac{\pi n}{2} \right) \nonumber \\
 && - \dfrac{Un}{2},
\end{eqnarray}
will take the following value at half-filling,
\begin{eqnarray}\label{eq:BALDA_n=1_pot}
\dfrac{\partial e_{\rm c}^{\rm BALDA}(n = 1)}{\partial n}
& = & 
  \left.\dfrac{\partial e_{\rm c}^{\rm BALDA}(n < 1)}{\partial n}
  \right|_{n=1^-} \nonumber \\
  & =&   - 
 2t
 \cos \left( \dfrac{\pi}{\beta(U/t)}\right) - \dfrac{U}{2}.
\end{eqnarray}
[Rather than being equal to 0 to fulfill Eq.~(\ref{eq:emb_pot_HF}).]
This choice makes the embedding potential smooth
in the range
$0 \leqslant n \leqslant 1$, instead of $0 \leqslant n < 1$.
We refer to this new functional with an overbar:
$\overline{\rm iBALDA}$. The
change between $\overline{\rm iBALDA}$
and iBALDA
operates only at half-filling and
is made clear by representing the correlation potential
in Fig.~\ref{fig:iBALDA}.
Note that in contrast to iBALDA,
the 
2L-BALDA correlation potential depicts no discontinuity at half-filling
(Fig.~9 of Ref.~\cite{senjean2018multiple}).

Why haven't we made this choice previously ? In SOET, convergence
issues may arise around half-filling when either the impurity {\it or}
the bath occupations come very close to 1. These issues
appear in iBALDA
if the BALDA correlation 
potential [Eq.~(\ref{eq:BALDA_n=1_pot})]
were used to solve the self-consistent SOET equation.
This problem is due to the presence of
the Mott metal-insulator transition at half-filling,
and also arises in KS-SOFT
(for inhomogeneous systems with BALDA)~\cite{lima2003density}.
In P-SOET, the situation is different.
The density is determined self-consistently from
a KS-SOFT calculation (which is always exact for a uniform system), while the embedded
problem is solved only once.
Therefore, Eq.~(\ref{eq:BALDA_n=1_pot})
can be used at half-filling in P-SOET without leading to
any convergence issue.
We show in Sec.~\ref{sec:results} that choosing the
$\overline{\rm iBALDA}$ embedding potential at half-filling
instead of the iBALDA one
improves the results drastically,
although the resulting impurity occupation is not exact anymore.
(In any case, the exact KS-SOFT occupations are used to compute the
per-site energy and the double occupation in P-SOET.)

\section{Computational details}\label{sec:comp_details}

Three main calculations are performed in P-SOET:
($i$) the KS-SOFT calculation of the physical problem to get 
$\Phi_{\rm KS}$ [Eq.~(\ref{eq:h_soft})], ($ii$) the Schmidt decomposition of $\Phi_{\rm KS}$
and ($iii$) solving the embedded problem [Eq.~(\ref{eq:H_emb})].
As the uniform one-dimensional Hubbard model 
is studied in this paper,
the first step simply consists in a mean-field calculation 
with no potential. Indeed, the KS potential is defined up to a constant,
and is also uniform for a uniform system.
The second step follows the implementation
detailed in Ref.~\cite{ayral2017dynamical}. Finally, the
embedded problem is solved either analytically for a 
single impurity site
(see Appendix~\ref{sec:dimer}) or
numerically for multiple impurities by using 
the Block code of density matrix renormalization group
(DMRG)~\cite{chan2002highly,chan2004algorithm,
ghosh2008orbital,sharma2012spin,olivares2015ab}.
The derivations of the SOET functionals can be found
in Ref.~\cite{senjean2018multiple}, and
are implemented in the source code for P-SOET 
which is freely available~\cite{psoet_code}.
In this work, the $L$-site uniform one-dimensional
Hubbard model is studied
with an even number $N$ of electrons.
Periodic ($\hat{c}_{L\sigma}=
\hat{c}_{0\sigma}$) and 
antiperiodic
($\hat{c}_{L\sigma}
=-\hat{c}_{0\sigma}$) boundary conditions have been used when   
$(N/2)~{\rm mod}~2 = 1$ (i.e., $N/2$ is an odd number)
and $(N/2)~{\rm mod}~2 = 0$ (i.e., $N/2$ is an even 
number), respectively.
Results are compared to the exact BA results~\cite{NoMott_Hubbardmodel,shiba1972magnetic,
Hubbardmodel}.
$L = 400$ sites are considered in this work and the hopping 
parameter has been set to $t = 1$ in all the calculations.

\section{Results}\label{sec:results}

After evaluating the
errors coming from the
projection (constructed from approximate 
single-particle bath states) onto the impurity model
subspace,
the per-site energy [Eq.~(\ref{eq:per_site_ener})] and
the double occupation [Eq.~(\ref{eq:dblocc_SOET})] are
calculated within P-SOET as described in Sec.~\ref{sec:PSOET}.
The focus will be first on the
half-filled case, followed by the hole-doping regime. Finally,
the density-driven Mott--Hubbard transition is investigated
and found to be properly described by P-SOET with 
a single impurity site, in contrast to
DMET~\cite{knizia2012density,bulik2014density} and
DMFT~\cite{capone2004cluster} for which multiple impurities
are required.

\subsection{Errors due to the projection}

Because the projector
is expressed in terms of 
approximate single-particle
states, P-SOET is no more in-principle exact
in contrast to SOET (providing that exact functionals are used).
It is therefore essential to evaluate the
errors due to this projection,
in addition to those introduced in the functionals.
Such an analysis is performed
by comparing the
impurity occupation and double
occupation
within SOET
and P-SOET
using the exact SOET embedding potential,
obtained by reverse engineering~\cite{senjean2017local} 
on the uniform $L=12$ sites 
Hubbard model with a single impurity site.
By construction,
the impurity occupation obtained in SOET by using
this potential is the exact
one~\cite{senjean2017local}.
However, the one obtained after projection is not
exact anymore
except in the particular case of half-filling,
as shown in Table~\ref{tab:LFpot}.
As readily seen in Table~\ref{tab:LFpot}, the impurity occupation
in P-SOET deviates from the exact occupation
as $U/t$ increases. Nevertheless the error remains relatively low, 
of the order of 
$10^{-2}$.

\begin{table*}
\begin{center}
\begin{tabularx}{1\textwidth}{lc *{6}{Y}}
\hline

& \multicolumn{1}{c}{$N=2$}
& \multicolumn{1}{c}{$N=4$}
 & \multicolumn{1}{c}{$N=6$}
 &  \multicolumn{1}{c}{$N=8$}
 & \multicolumn{1}{c}{$N=10$} 
  &  \multicolumn{1}{c}{$N=12$} \\
  \hline
Exact & 0.16667 & 0.33333 & 0.5 & 0.66667 & 0.83333 & 1\\
$U=t$    &0.16559 &0.33174 &0.49883 &0.66618 &0.83329 &1\\
$U=2t$  &0.16377 &0.32854 &0.49613 &0.66488 &0.83305 &1\\
$U=3t$  &0.16209 &0.32520 &0.49296 &0.66307 &0.83251 &1\\
$U=4t$  &0.16068 &0.32223 &0.48990 &0.66104 &0.83167 &1\\ 
$U=5t$  &0.15954 &0.31972 &0.48719 &0.65904 &0.83057 &1\\
$U=6t$  &0.15861 &0.31765 &0.48489 &0.65718 &0.82930 &1\\
$U=7t$  &0.15784 &0.31593 &0.48296 &0.65553 &0.82796 &1 \\
$U=8t$  &0.15720 &0.31449 &0.48134 &0.65409 &0.82663 &1\\
$U=9t$  &0.15666 &0.31327 &0.47997 &0.65284 &0.82537 &1\\
$U=10t$ &0.15621 &0.31224 &0.47881 &0.65177 &0.82421 &1\\
$U=100t~~~~~$&0.15141 &0.30184 &0.46785 &0.63440 &0.81267 &1 \\
\hline
\end{tabularx}
\caption{Impurity occupation obtained by solving the embedded 
problem in P-SOET. The exact SOET
embedding potential for a uniform $L=12$ sites 
Hubbard model with one impurity site has been used.}
\label{tab:LFpot}
\end{center}
\end{table*}

\begin{figure}
\centering
\resizebox{\columnwidth}{!}{
\includegraphics[scale=1]{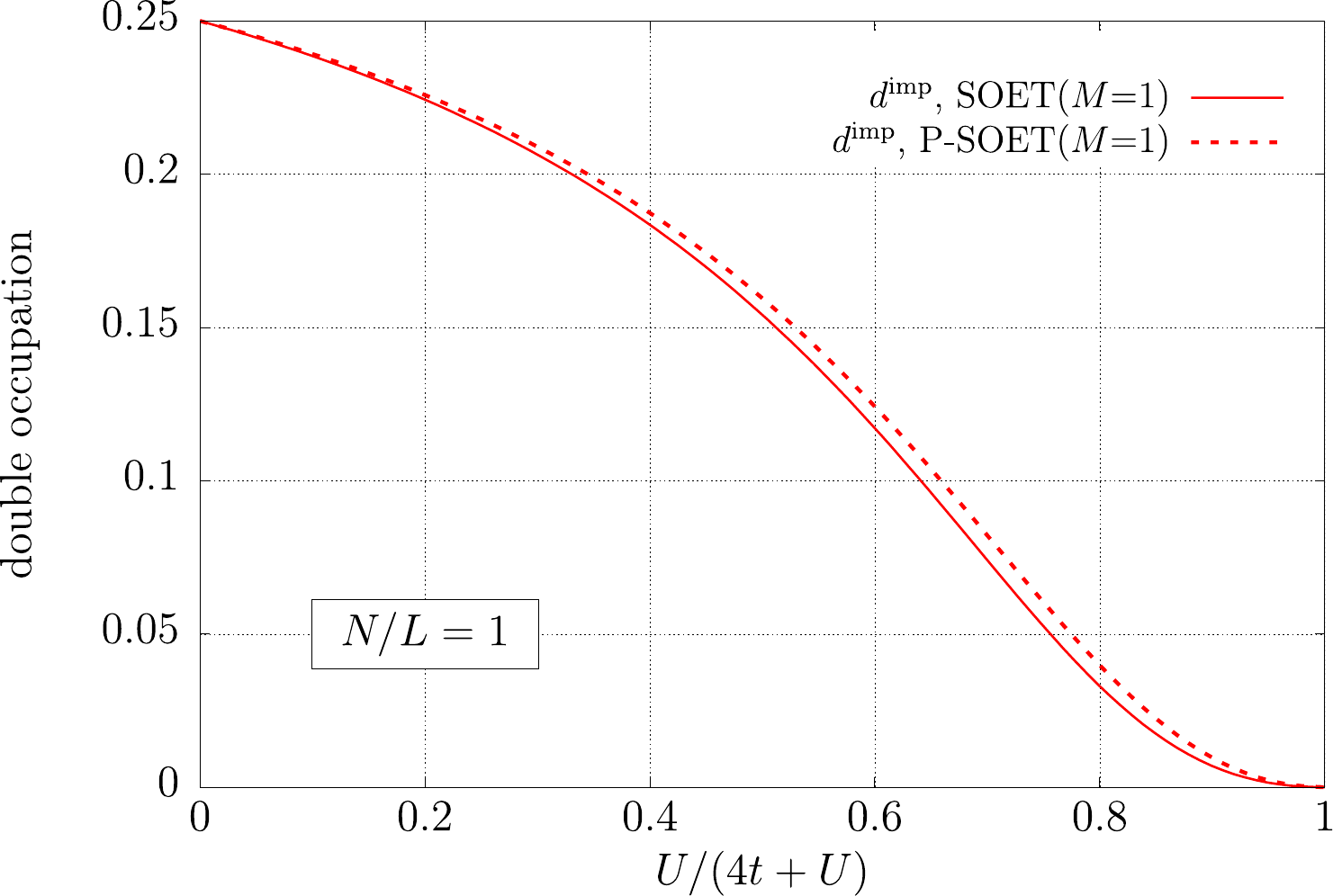}
}
\caption{Impurity double occupation
$d^{\rm imp} = \langle 
\hat{n}_{0\uparrow}\hat{n}_{0\downarrow} \rangle_{\Psi_M}$
as a function of correlation strength.
The wavefunction $\Psi_M$ is obtained by solving the SOET 
Hamiltonian [Eq.~(\ref{eq:self-consistent-SOET_new})] ($\Psi^{\rm imp}_M$, full lines),
and the embedded problem in P-SOET [Eq.~(\ref{eq:H_emb})] 
($\Psi^{\rm emb}_M$, dashed lines). The uniform and 
half-filled Hubbard model with $L=12$ and a single
impurity site is considered.
In both cases, the exact embedding potential
in Eq.~(\ref{eq:emb_pot_HF}) has been used.}
\label{fig:dimp}
\end{figure}

Turning to the impurity double occupation, let us
consider the half-filled case where the exact potential does indeed
lead to the exact impurity occupation, even in P-SOET.
In Fig.~\ref{fig:dimp}, the impurity double occupation is
shown for both SOET and P-SOET with respect to $U/(U + 4t)$.
Note that the range $0 \leqslant U/(U + 4t) \leqslant 1$ 
covers the entire correlation regime,
from the weakly correlated one $U < 4t$ [$U/(U + 4t) < 1/2$],
to the strongly correlated one $U > 4t$ [$U/(U + 4t) > 1/2$]. The
noninteracting and atomic limits are given by $U = 0$ [$U/(U + 4t) = 0$] and $t = 0$ [$U/(U + 4t) = 1$], respectively.
We observe a very small change between the impurity double occupations in Fig.~\ref{fig:dimp}, and 
conclude that the projection does not lead to substantial
errors. Therefore, the errors will almost entirely
be due 
to the use of approximate functionals.

The iBALDA and 2L-BALDA results within P-SOET
are then not expected to
differ much from the ones obtained with 
SOET in Ref.~\cite{senjean2018multiple},
and are mostly plotted for comparison with the novel
$\overline{\rm iBALDA}(M)$ approximation introduced in Sec.~\ref{sec:approx}.

\subsection{Half-filled case}

\begin{figure}
\centering
\resizebox{\columnwidth}{!}{
\includegraphics[scale=1]{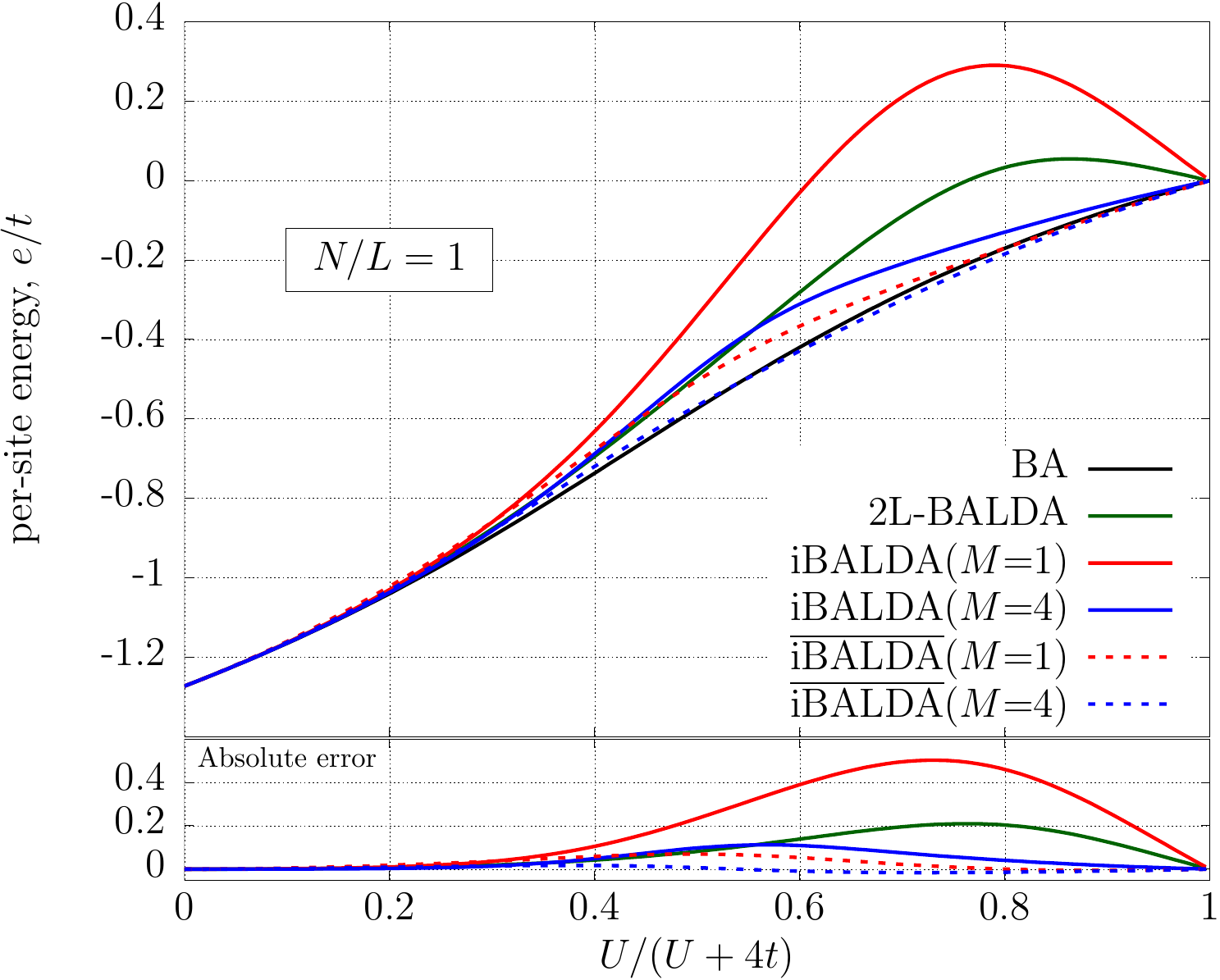}
}
\caption{Per-site energy [Eq.~(\ref{eq:per_site_ener})]
as a function of correlation strength
for 
different approximations in P-SOET. 
The uniform and half-filled $L=400$ site Hubbard is considered.
The absolute error with respect to BA is plotted in the bottom panel.}
\label{fig:persite_halffilling}
\end{figure}

Let us first focus on the half-filled Hubbard model, 
known to be
a Mott insulator for any $U > 0$.
Fig.~\ref{fig:persite_halffilling} shows the
per-site energy obtained
within P-SOET for different approximations.
First of all, all the approximations are exact
in the noninteracting and atomic limits.
As already discussed in Ref.~\cite{senjean2018multiple}, it is clear
that the iBALDA($M$=1) is not accurate enough at half-filling
due to the absence of bath correlation functional, above all
in the moderate and strong correlation regimes.
Considering a non-zero bath correlation functional
like in 2L-BALDA improves over iBALDA($M$=1). With a single impurity 
site, 2L-BALDA gives similar results than iBALDA($M$=4) in the 
weak and moderate correlation regime but is less accurate in the
strongly correlated one.
Still considering a single impurity, 
$\overline{\rm iBALDA}$($M$=1) is
surprisingly significantly more accurate than iBALDA($M$=1) and 2L-BALDA
for $U \geqslant 4t$. By increasing the number of impurities,
$\overline{\rm iBALDA}$($M$=4) is now almost 
on top of the exact BA per-site energy. 
In comparison with DMET
(Fig.~2 of 
Ref.~\cite{bulik2014density}),
$\overline{\rm iBALDA}$
gives a better per-site energy
than the so-called ``NI'', and is even competitive with the so-called
``NI$_F$'', which are nothing but the noninteracting bath formulations 
of DMET where the matching condition is performed 
on the 1RDM of the fragment and on the full 1RDM of the
fragment+bath, respectively~\cite{bulik2014density}.
We refer the reader to Ref.~\cite{bulik2014density} 
for more details about those two versions of DMET.

\begin{figure}
\centering
\resizebox{\columnwidth}{!}{
\includegraphics[scale=1]{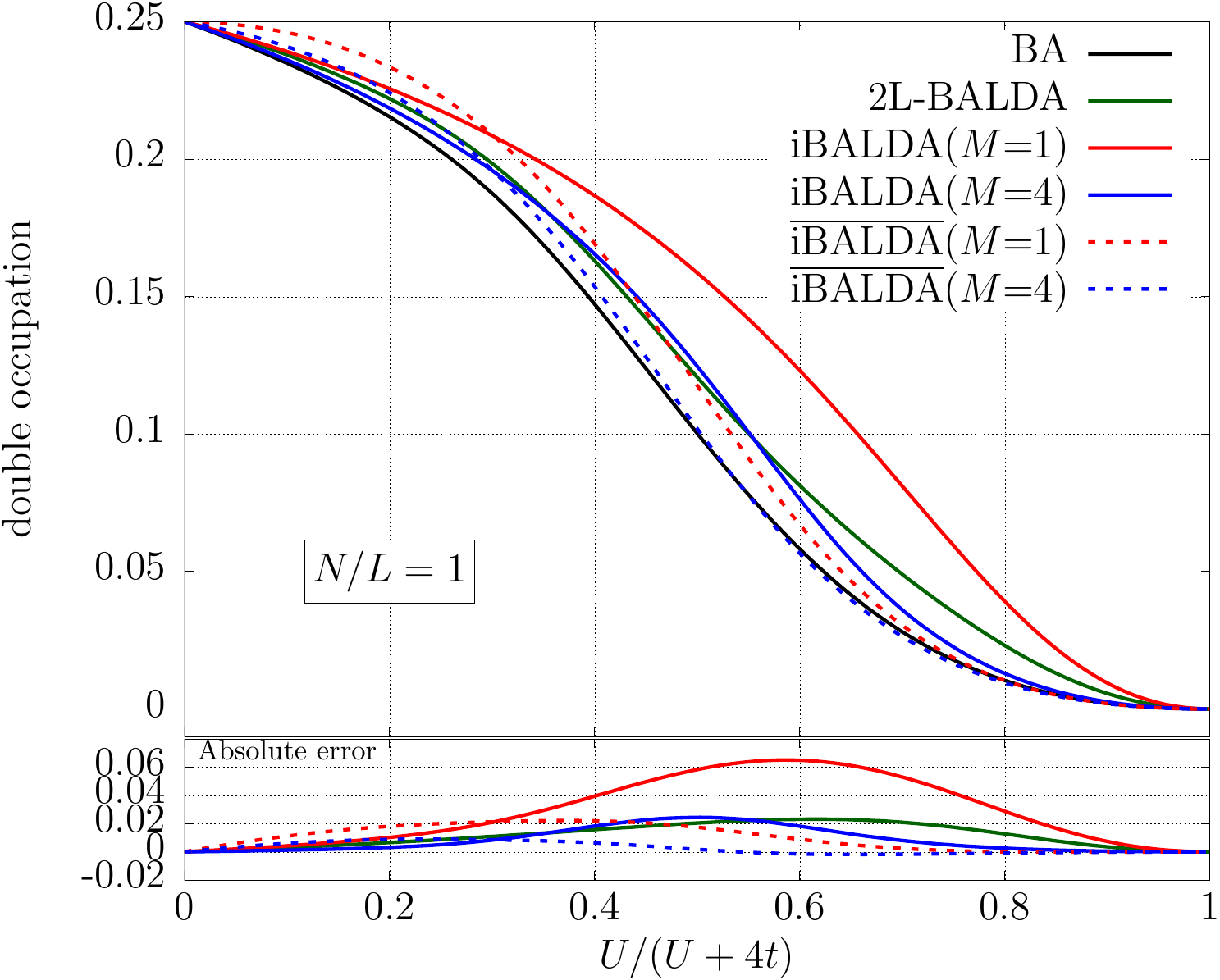}
}
\caption{Double occupation [Eq.~(\ref{eq:dblocc_SOET})]
as a function of correlation strength
for 
different approximations in P-SOET. 
The uniform and half-filled $L=400$ site Hubbard is considered.
The absolute error with respect to BA is plotted in the bottom panel.}
\label{fig:dblocc_halffilling}
\end{figure}

Let us now turn to the double occupation in 
Fig.~\ref{fig:dblocc_halffilling}. As expected,
the double occupation within iBALDA and 2L-BALDA
are similar to the one obtained in
SOET~\cite{senjean2018multiple}, demonstrating again that errors due 
to the projection are negligible.
Interestingly, the double occupation obtained
within iBALDA($M$=1) is identical to the one of
DMET
for a single impurity site~\cite{knizia2012density,bulik2014density}.
This can be explained as follows: for a single impurity site at half-filling
the iBALDA embedding potential returns the exact density, 
exactly like the numerically
optimized correlation potential of DMET. Therefore,
we see no reason why the embedded wavefunction
of both theories should be different in this case, thus leading to
the same impurity
double occupation.
However, the method
overestimates the exact double occupation.
Just like the per-site energy,
2L-BALDA improves
over iBALDA($M$=1) in the entire correlation regime. 
An even better
double occupation is obtained with iBALDA($M$=4) especially in the
strongly correlated regime, but this is at the expense of a much higher
computational cost. Indeed, a single impurity can be solved analytically
(see Appendix~\ref{sec:dimer}) while four impurities require to solve
a fragment+bath system of eight sites (or 16 spin orbitals)
and eight electrons with a high-level wavefunction method.
Turning to
$\overline{\rm iBALDA}$($M$=1), the double occupation
is worse than iBALDA($M$=1) for $U \leqslant 1.8t$, but becomes
rapidly more accurate than 2L-BALDA or even iBALDA($M$=4).
With four impurity sites,
$\overline{\rm iBALDA}$($M$=4) is also
not highly accurate in the very weakly correlated regime, but
tends towards the exact double occupation otherwise.

Finally,
the fact that the bare 
(i.e. without density functional contributions) impurity double occupation
changes that much between
iBALDA and $\overline{\rm iBALDA}$ means
that the embedding
potential has a strong influence on the
embedded wavefunction.
Although the embedding potential
in $\overline{\rm iBALDA}$ is no more exact at half-filling
for a finite
system and
does not lead to the
exact density (as discussed in Sec.~\ref{sec:approx}), 
it may take into account the
so-called exchange-correlation derivative-discontinuity responsible
for the description of the Mott insulator. This derivative-discontinuity
has been discussed
by Capelle {\it et al.} for the
BALDA functional~\cite{lima2002density,capelle2003density}
and in Ref.~\cite{senjean2018site} for SOET functionals.
This could explain
why $\overline{\rm iBALDA}$
is the most accurate approximation
at half-filling (in the moderate and strong correlation regimes).

\subsection{Hole-doped regime}

Let us now investigate the hole-doped region 
($0.4 \leqslant N/L \leqslant 1$) of the uniform 
Hubbard model for moderate ($U = 4t$) and strong
($U = 8t$) correlation strengths.
Note that in 
previous works~\cite{senjean2018site,senjean2018multiple}, 
we were limited
to $L = 32$ sites for computational cost reason.
Within P-SOET, $L=400$ sites is easily tractable as the
most costly part now consists in solving an embedded problem
of 2$L_F$ sites only. As a consequence,
we can get closer to the thermodynamic limit.
Because iBALDA and $\overline{\rm iBALDA}$ differ
at half-filling only, they will give exactly the same result
in the density domain $0.4 \leqslant N/L < 1$. 
To have smooth potentials
in the range $0.4 \leqslant N/L \leqslant 1$, $\overline{\rm iBALDA}$
is used instead of
iBALDA (see Fig.~\ref{fig:iBALDA}).

\begin{figure}
\centering
\resizebox{\columnwidth}{!}{
\includegraphics[scale=1]{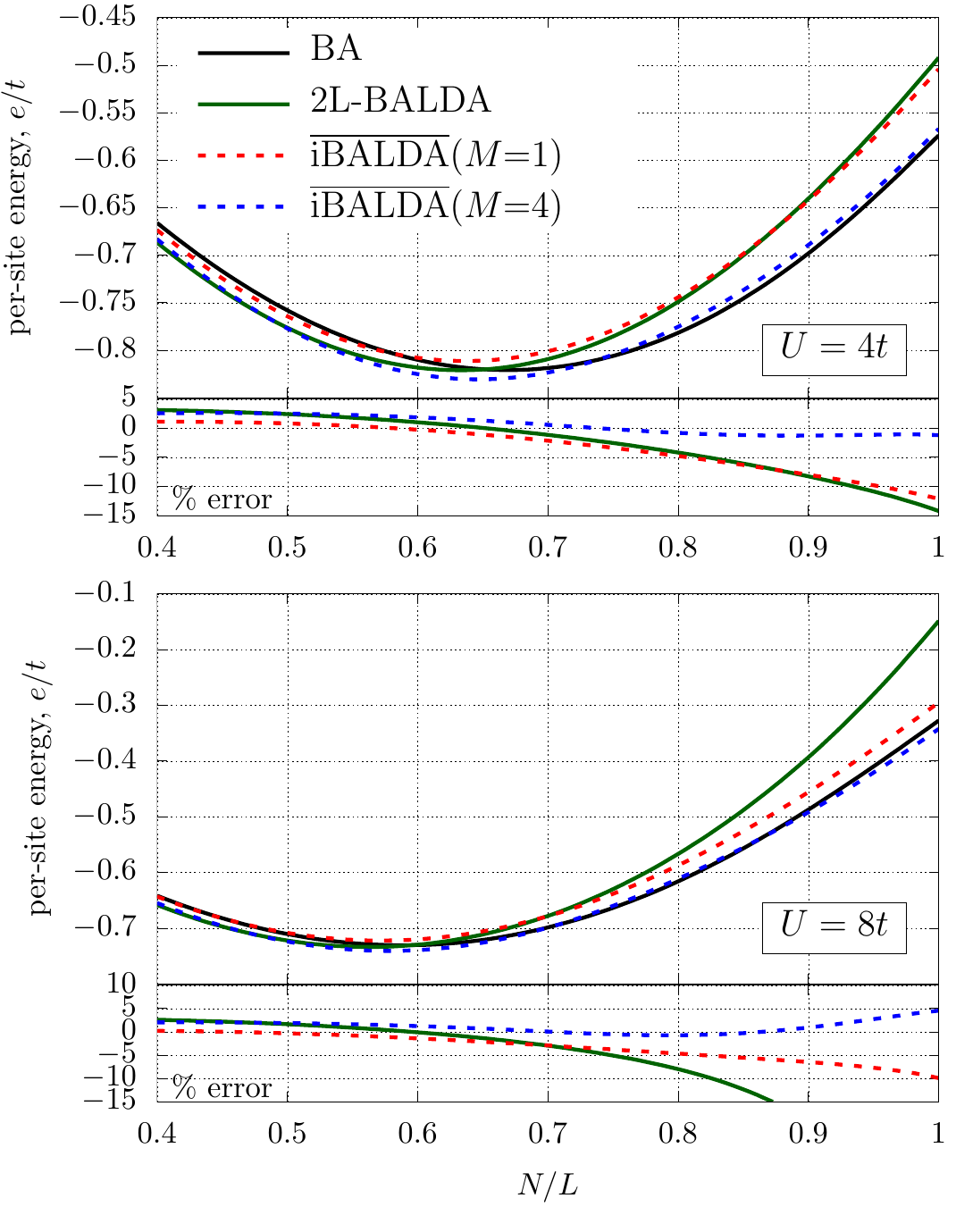}
}
\caption{Per-site energy as a function of the density filling
for different approximations in P-SOET. The
$L = 400$ sites Hubbard model with $U = 4t$ (top panel) and 
$U = 8t$ (bottom panel) is considered.
The relative error with respect to BA is plotted in the bottom of each panel.}
\label{fig:persite}
\end{figure}

In the case of $U = 4t$ (top panel of Fig.~\ref{fig:persite}),
2L-BALDA and $\overline{\rm iBALDA}$($M$=1) 
are already fairly accurate and almost indistinguishable from
each other. 
Nevertheless, they both 
tends to overestimate the per-site energy closer they get to 
half-filling. Increasing the number of impurities
reduces the error considerably, 
and $\overline{\rm iBALDA}$($M$=4) 
becomes almost exact for $N/L \geqslant 0.7$.
For stronger correlation strength (bottom panel of
Fig.~\ref{fig:persite}),
$\overline{\rm iBALDA}$($M$=1) is now
much more accurate. By comparison with Fig.~3 of
Ref.~\cite{bulik2014density}, the single-impurity 
$\overline{\rm iBALDA}$($M$=1)
gives a better per-site energy than
NI and NI$_F$ with two impurities. It
is even comparable to DET and DMET
in the broken
spin symmetry formalism, also with two impurities~\cite{bulik2014density}.
2L-BALDA still overestimates the per-site energy close to half-filling,
while $\overline{\rm iBALDA}$($M$=4) is again almost exact.

\begin{figure}
\centering
\resizebox{\columnwidth}{!}{
\includegraphics[scale=1]{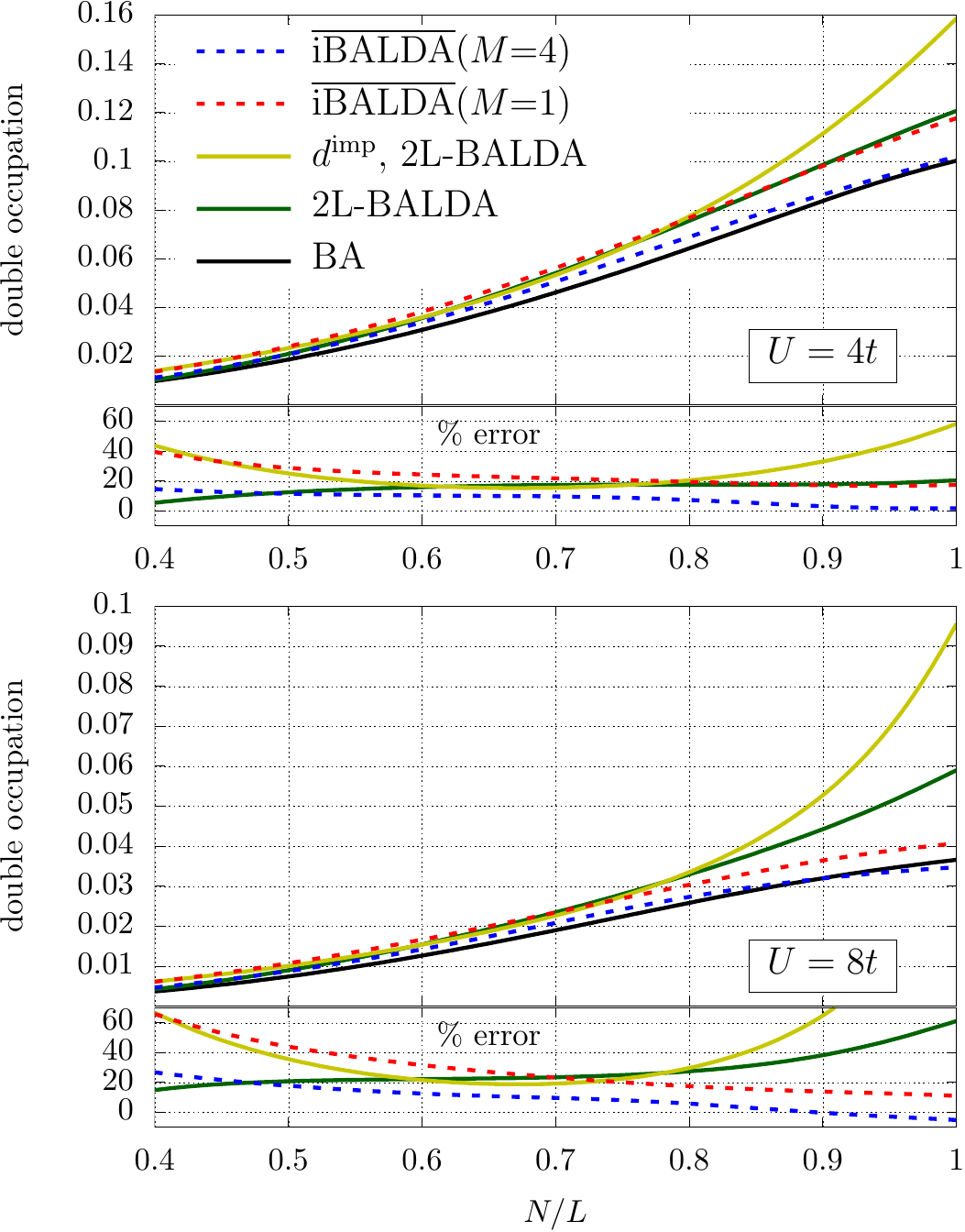}
}
\caption{Double occupation as a function of the density filling
for different approximations in P-SOET. The
$L = 400$ sites Hubbard model with $U = 4t$ (top panel) and 
$U = 8t$ (bottom panel) is considered.
The relative error with respect to BA is plotted in the bottom of each panel.
}
\label{fig:dblocc}
\end{figure}

Turning to the double occupation in Fig.~\ref{fig:dblocc},
one can see that the bare impurity double occupation $d^{\rm imp}$
obtained with 2L-BALDA
overestimates the exact double occupation considerably
around half-filling. The physical double occupation 
within 2L-BALDA is more accurate as expected.
The analysis of Fig.~\ref{fig:dblocc} is then similar to
Fig.~\ref{fig:persite}, due to the relation between 
the double occupation and
the per-site energy [Eq.~(\ref{eq:per_site_dblocc})]. (The term in 
square brackets in Eq.~(\ref{eq:per_site_dblocc})
is approximated by BALDA, which is accurate in all regimes
except the very weakly correlated one~\cite{senjean2018site}, 
not studied here).
Therefore the double occupation
within $\overline{\rm iBALDA}$($M$=1) is also 
very accurate, especially for $U = 8t$. Increasing the number of
impurities does improve the result slightly, 
but might not be worth it here considering its higher computational
cost.

\subsection{Density-driven Mott--Hubbard transition}\label{sec:transition}

\begin{figure}
\centering
\resizebox{\columnwidth}{!}{
\includegraphics[scale=1]{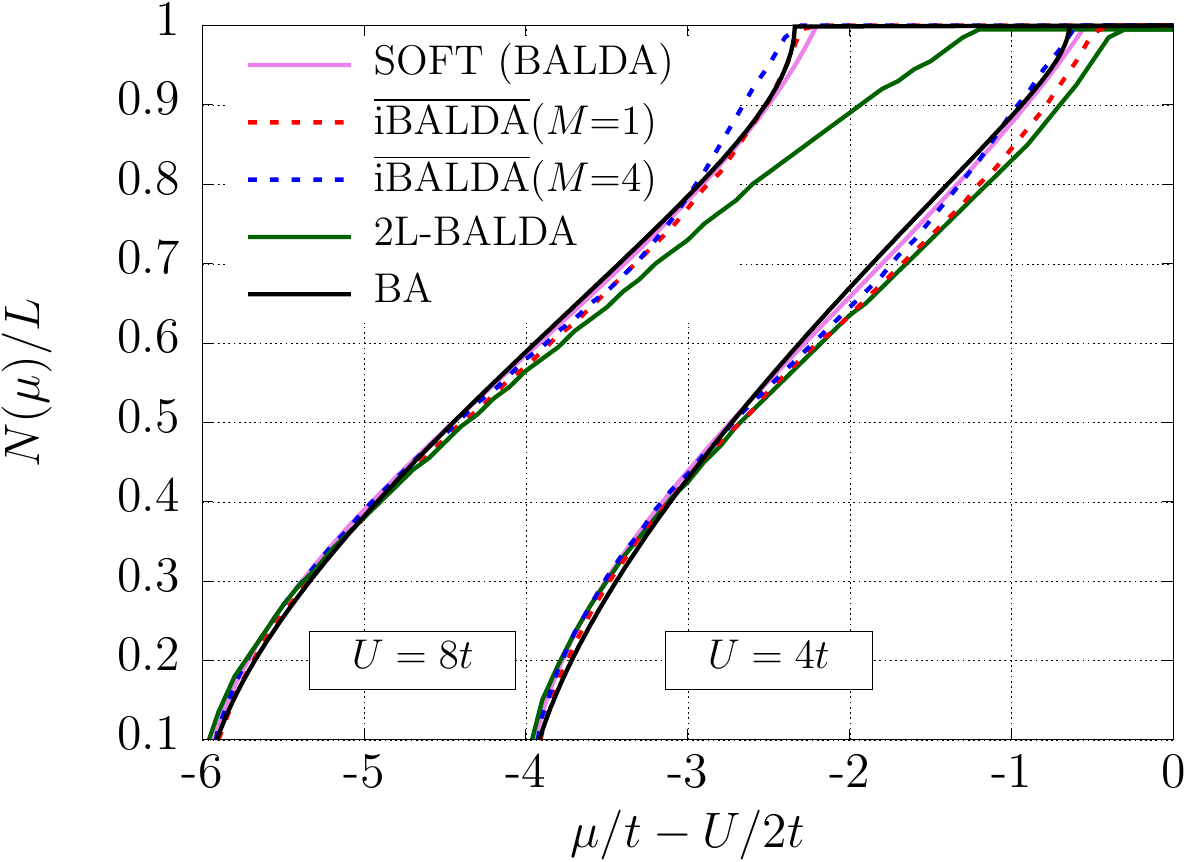}
}
\caption{Lattice filling as a function of the chemical
potential determined according to Eq.~(\ref{eq:Nmu})
for different approximations within P-SOET,
and for the BALDA within SOFT.
The curves correspond to $U= 8t$ on the left side and
$U=4t$ on the right side.}
\label{fig:transition}
\end{figure}

The most interesting behavior of the one-dimensional
Hubbard model is certainly 
the paramagnetic density-driven Mott--Hubbard transition.
In one dimension, the Hubbard model describes a paramagnetic
metallic state,
except at half-filling where it becomes a paramagnetic Mott-insulator
(for any $U > 0$) manifested by the opening of a charge gap. 
Therefore, a density-driven Mott metal-insulator
transition arises when
passing from the infinitesimally hole-doped case ($N/L = 1-\delta$, 
with $\delta$ infinitesimal and strictly positive)
to the half-filled case ($N/L = 1$). Such a transition,
which is not a result of antiferromagnetism,
is detected by vanishing compressibility
(or charge susceptibility 
$\partial N(\mu) / \partial \mu$)~\cite{knizia2012density}.
It can be observed
by plotting $N(\mu)$ with respect to
$\mu$, where $N(\mu)$ is the electron number
associated with the chemical potential
$\mu$ and is obtained by minimizing
the grand canonical energy of the following Hamiltonian:
\begin{eqnarray}\label{eq:H-H0-muN}
\hat{H} = \hat{H}_0 - \mu \hat{N},
\end{eqnarray}
with $\hat{H}_0$ being the Hubbard Hamiltonian and
$\hat{N}$ the counting operator.
Both paramagnetic
single-site DMET~\cite{knizia2012density}
and paramagnetic
single-site DMFT~\cite{capone2004cluster}
do not feature the opening of the charge gap in the
one-dimensional Hubbard model.
Interestingly, matching the 1RDM of the fragment only in the
non-interacting bath DMET (NI in Figs. 4 and 5 of Ref.~\cite{bulik2014density})
does not predict a gap even by considering multiple impurity sites,
while fitting the entire fragment+bath 1RDM
does~\cite{knizia2012density}.
To observe this transition,
it comes from Eq.~(\ref{eq:H-H0-muN})
that we have to perform the
minimization of the per-site grand canonical energy:
\begin{eqnarray}\label{eq:Nmu}
N(\mu) = \argmin_{N}\Big \lbrace e(N/L) - \mu N/L\Big \rbrace,
\end{eqnarray}
where $e(N/L)$ is the per-site energy
obtained within P-SOET for a system
of $N$ electrons and $L$ sites. The resulting number of electrons
$N(\mu)$ with respect to the chemical potential is given
in Fig.~\ref{fig:transition}.

As readily seen in Fig.~\ref{fig:transition}, $\overline{\rm iBALDA}$
does predict the opening of the charge gap with high accuracy,
for both $U=4t$ and $U=8t$.
Most impressively, only a single impurity site
is sufficient to describe both the tendency of
the exact curve as well as the actual position of the Mott transition.
Increasing the number of impurities does not lead to any
particular change.
The transition seems to be also well reproduced within 2L-BALDA
for $U=4t$, but the vanishing compressibility is manifested
at $N(\mu) = 398$ instead of the half-filled value $N(\mu) = 400$.
The same error arises for $U=8t$. Besides, the position 
of the Mott transition is not anymore correct and the value
of the chemical potential is significantly overestimated compared
to the BA results.

This striking result is a first for this type of embedding methods.
It can be attributed to the additional use of 
accurate density functionals 
(BALDA here). Indeed, we can see that the BALDA curve 
within SOFT
is almost on top of the BA results.
The fact that BALDA is able 
to describe the opening of the charge gap (driven
by non-local correlation effects in 1D~\cite{capone2004cluster})
is due to the presence of a derivative discontinuity 
in the functional.
This so-called
``ultranonlocality'' of BALDA
(as opposed to the LDA in the continuum) has been
extensively discussed by Ying {\it et al.} in Ref.~\cite{ying2014solving}.
To further stress on the
importance of having a derivative discontinuity,
we have proved in Appendix D of Ref.~\cite{senjean2018site}
that
the impurity correlation functional $E_{\rm c}^{\rm imp}(\bfn)$ 
and the per-site correlation functional $e_{\rm c}(n)$ should manifest 
a derivative discontinuity
at half-filling, while the bath
correlation functional $\overline{E}_{\rm c}^{\rm bath}(\bfn)$
should not. 
Even though the proof was derived in the atomic limit, 
we still think that
the derivative discontinuity can be important 
for finite interaction strength.
This condition is fulfilled by
$\overline{\rm iBALDA}$ but not by 2L-BALDA
(see the impurity correlation potential in 
Fig.~9 of Ref.~\cite{senjean2018multiple}), thus
justifying why the Mott transition is better described within
$\overline{\rm iBALDA}$.
In DMFT, non-local effects
can be described by considering a cluster
rather than a single site~\cite{capone2004cluster}.
A new formulation of SOET using the Green's function formalism
could provide a link between
density functional contributions and the self-energy of the cluster,
thus making a clearer connection between SOET and (cluster) DMFT.
This is left for future work.

\section{Conclusions and perspectives}\label{sec:conclu}

In this work, a new method so-called P-SOET
has been derived from SOET
by using the Schmidt decomposition.
P-SOET is shown to give
similar results as SOET
but at a drastically reduced computational cost,
thus allowing for calculations much closer to the thermodynamic limit.
The use of the Schmidt decomposition makes P-SOET
competitive with DMET~\cite{knizia2012density} and
DET~\cite{bulik2014density},
with additional density functional contributions.
As such, the embedding potential in P-SOET
is analytical and is expressed as an energy derivative
functional of the density, instead of being numerically optimized like
in DMET.
Accurate per-site energies
and double occupations in the entire correlation
regime and density domain have been obtained
on the one-dimensional Hubbard model with 400 sites.
As an important result, 
the density-driven Mott--Hubbard transition has
been accurately predicted with a single impurity site.
As far as we know, this is
a first for this kind of embedding methods.

Note that SOET or
P-SOET are not limited to the uniform 
one-dimensional Hubbard model
and several extensions can be considered.
First of all, (P-)SOET is straightforwardly
applicable to higher-dimensional systems
as the theory remains the same. The only requirement is that
new appropriate functionals are needed. A recent work by
Vilela {\it et al.}~\cite{vilela2019approximate}
could be used to derive a functional for the 
two- and three-dimensional Hubbard models from 
dimensional scaling of the BALDA functional.
The description of heterogeneous systems
can also be done by dividing the full system into
multiple fragments. A global chemical potential
would then be numerically optimized
to ensure that the occupations of the fragments sum
up to the total number of electrons, in the line of DMET.
Magnetic and spin-dependent phenomena can also 
be addressed by
including dependence on spin~\cite{franca2012simple}
and on the current density~\cite{akande2012persistent}
in the functional.
The treatment of finite temperature effects
would require a state-average
theory and a temperature-dependent functional. While such functional 
could be relatively easily developed
(see for instance Ref.~\cite{xianlong2012lattice}), 
and the state-average extension comes naturally within SOET, 
P-SOET relies on the Schmidt decomposition
which is state-dependent. 
The same problem arises in DMET. 
Note that an extension to excited states within DMET has been 
recently addressed by Tran {\it et al.}~\cite{tran2019using}. Their
work could be used for similar developments in P-SOET.
Turning to transport properties and the description of systems out of 
equilibrium, a real-time extension of P-SOET can be derived
in analogy with real-time DMET~\cite{kretchmer2018real}, together 
with time-dependent functionals that could be
derived from earlier works on time-dependent lattice 
DFT~\cite{verdozzi2008time,kurth2010dynamical,karlsson2011time,
kurth2017transport,carrascal2018linear,kurth2018time}.
Alternatively, a frequency-dependent formulation of SOET could be 
derived using the
Green's function formalism.
Finally, the extension
to realistic systems like molecules
(as done in DMFT~\cite{zgid2011DMFTquantum}, 
DMET~\cite{knizia2013density,wouters2016practical}, and
SEET~\cite{lan2015communication})
is maybe the most important one.
Indeed,
embedding approaches are appropriate to treat
large systems in quantum chemistry
due to their good balance between accuracy
and computational cost
(see Refs.~\cite{wesolowski2015frozen,lee2019projection} and references therein).
By decomposing the full system into
subsystems, they are also promising candidates
for solving classically intractable
chemistry problems on
near-term quantum
devices~\cite{bauer2016hybrid,rubin2016hybrid,
reiher2017elucidating,
yamazaki2018towards}
(see Ref.~\cite{mcardle2018quantum} for a 
review on quantum computational chemistry).
By combining WFT
and DFT,
P-SOET would then be
an efficient and free from double counting
low cost embedding method able to treat
both dynamical and static correlation effects 
of large chemical systems.
Unfortunately, this
extension is not straightforward as
it faces fundamental issues like
the
dependence of the functionals
on the molecular orbital basis~\cite{senjean2018thesis}.
Nevertheless, recent works on extending lattice DFT 
to quantum chemistry
using localized orbitals~\cite{coe2019lattice} or also
the domain separation in DFT~\cite{mosquera2019domain}
could be helpful to generalize (P-)SOET to quantum chemistry.
To get rid of the dependence on the molecular orbital basis,
one could also work with the 
one-particle reduced density matrix as a variable instead of the
occupation number only.
P-SOET
sheds a new light on the treatment of
strongly correlated systems,
and could progress in any of these aforementioned directions,
which are under investigation.
The ideas highlighted in this paper
could hopefully inspire other works in the field.

\section{Acknowledgments}

BS is grateful to Emmanuel Fromager and Matthieu Sauban\`ere
for several fruitful discussions and meaningful comments 
on the article.
BS thanks Matthieu Sauban\`ere
for providing him the Bethe Ansatz program.
The DMRG calculations 
have been performed on the High-Performance 
Computing Center of the University of Strasbourg.

\newcommand{\Aa}[0]{Aa}

\appendix

\section{Analytical solution for the Anderson dimer}\label{sec:dimer}

The ground-state of the Anderson dimer 
is obtained
by solving
\begin{eqnarray}\label{eq:dimer_eq}
\hat{H}^{\rm imp} \vert \Psi^{\rm imp}\rangle  = \mathcal{E}^{\rm imp} \vert \Psi^{\rm imp} \rangle,
\end{eqnarray}
where
\begin{eqnarray}\label{eq:Hdimer}
\hat{H}^{\rm imp} = -t \sum_{\sigma} \left(\hat{c}_{0\sigma}^\dagger \hat{c}_{1\sigma}
+ {\rm H.c.} \right)
+U \hat{n}_{0\uparrow} \hat{n}_{0\downarrow} + \dfrac{\Delta v}{2} (\hat{n}_1 - \hat{n}_0)\nonumber \\
\end{eqnarray}
and $\Delta v = v_1 - v_0$.
For two electrons,
$\mathcal{E}^{\rm imp}$ has been
shown to be related to the fully-interacting
two-electron Hubbard dimer by a simple scaling and shifting 
relation~\cite{senjean2017local}:
\begin{eqnarray}\label{eq:shift_rel}
\mathcal{E}^{\rm imp}(U,t,\Delta v) = E(U/2,t,\Delta v - U/2).
\end{eqnarray}
$E(U,t,\Delta v)$ is the ground-state singlet energy, solution
of
\begin{eqnarray}\label{eq:solving_dimer}
- 4t^2U + \left(4t^2 - U^2 + \Delta v ^2 \right)E + 2UE^2 = E^3,
\end{eqnarray}
and can be expressed analytically
as follows~\cite{carrascal2015hubbard}:
\begin{eqnarray}\label{eq:dimer_solution}
E(u,t,\Delta v) = \dfrac{4t}{3}(u-w\sin(\dfrac{\pi}{6}+\theta)),
\end{eqnarray}
where
\begin{eqnarray}
u &=& U/(2t), \nonumber \\
w &=& \sqrt{3(1+\nu^2) + u^2}, \nonumber \\
\nu &=& \Delta v /(2t),\nonumber \\
\cos(3\theta) &=& (9(\nu^2 - 1/2) - u^2)u/w^3. \nonumber
\end{eqnarray}

According to the Hellmann--Feynman theorem, it follows
from
Eqs.~(\ref{eq:dimer_eq}) and (\ref{eq:Hdimer}) that the impurity occupation reads
\begin{eqnarray}\label{eq:n0}
n_0 &=&1 -\dfrac{\partial \mathcal{E}^{\rm imp}(U,t,\Delta v)}{\partial \Delta v}, \nonumber \\
& =&  1 - \left.\dfrac{\partial E(U,t,\Delta v)}{\partial \Delta v}\right|_{U=\overline{U},\Delta v = \overline{\Delta v}}.
\end{eqnarray}
where $\overline{U} = U/2$ and 
$\overline{\Delta v} = \Delta v - U/2$ are coming
from the scaling and shifting relation in Eq.~(\ref{eq:shift_rel}).
Similarly,
the off-diagonal element of the 1RDM of the Anderson dimer is given by
\begin{eqnarray}\label{eq:gamma_01}
\gamma_{01} & = & -\dfrac{1}{2}\dfrac{\partial \mathcal{E}^{\rm imp}(U,t,\Delta v)}{\partial t} \nonumber \\
& = & - \dfrac{1}{2} \left.\dfrac{\partial E(U,t,\Delta v)}{\partial t}
\right|_{U=\overline{U},\Delta v = \overline{\Delta v}}
\end{eqnarray}
as well as the impurity double occupation,
\begin{eqnarray}\label{eq:dimp_dimer}
d^{\rm imp} & = & \dfrac{\partial \mathcal{E}^{\rm imp}(U,t,\Delta v)}{\partial U} \nonumber \\
& = & \dfrac{\partial U/2}{\partial U}\left.\dfrac{\partial E(U,t,\Delta v)}{\partial U}\right|_{U=\overline{U},\Delta v = \overline{\Delta v}} \nonumber \\
& & + \dfrac{\partial \overline{\Delta v}}{\partial U} \left.\dfrac{\partial E(U,t,\Delta v)}{\partial \Delta v}\right|_{U=\overline{U},\Delta v = \overline{\Delta v}},\nonumber \\
& = & \dfrac{1}{2}\left( \left.\dfrac{\partial E(U,t,\Delta v)}{\partial U}\right|_{U=\overline{U},\Delta v = \overline{\Delta v}} - (1 - n_0)\right). \nonumber \\
\end{eqnarray}
The derivatives of the ground-state energy
of the two-electron Hubbard dimer can be found by differentiating
Eq.~(\ref{eq:solving_dimer}),
thus leading to
\begin{eqnarray}
\dfrac{\partial E}{\partial \Delta v} &=& \dfrac{2\Delta v E}{3E^2 - 4UE + U^2 - 4t^2 - \Delta v^2},\nonumber \\
\dfrac{\partial E}{\partial t} &=& \dfrac{8t(E - U)}{3E^2 - 4UE + U^2 - 4t^2 - \Delta v^2}, \\
\dfrac{\partial E}{\partial U} &=& \dfrac{2E(E-U) - 4t^2}{3E^2 - 4UE + U^2 - 4t^2 - \Delta v^2},\nonumber
\end{eqnarray}
where $E$ is given by Eq.~(\ref{eq:dimer_solution}).

In P-SOET with a single impurity site, the embedded Hamiltonian
reduces to the Hamiltonian of 
an Anderson dimer with two electrons that reads:
\begin{eqnarray}
\hat{H}^{\rm emb} = \sum_{ij=0,\sigma}^1 h^{\rm emb}_{ij}
\left(
\hat{c}_{i\sigma}^\dagger \hat{c}_{j\sigma} + {\rm H.c.} \right)
+ U \hat{n}_{0\uparrow} \hat{n}_{0\downarrow}.\nonumber \\
\end{eqnarray}
The impurity occupation, the off-diagonal of the 1RDM and the 
impurity double occupation
are calculating from Eqs.~(\ref{eq:n0}), (\ref{eq:gamma_01}) 
and (\ref{eq:dimp_dimer})
by replacing $t$ and $\Delta v$ by
and $t = -h^{\rm emb}_{01} = - h^{\rm emb}_{10} > 0$
and
$\Delta v = h^{\rm emb}_{11} - h^{\rm emb}_{00}$.

\section{Self-consistency in P-SOET}\label{sec:self-cons}

\subsection{Self-consistent procedure}

SOET is a variational theory with respect to the density~\cite{fromager2015exact,
senjean2017local,senjean2018site,senjean2018multiple}.
This is different in P-SOET, which is split in two different part:
the KS problem [Eq.~(\ref{eq:h_soft})]
and the embedded problem [Eq.~(\ref{eq:H_emb})]
obtained by
projection of the effective Hamiltonian [Eq.~(\ref{eq:h_eff})].
The solution of the KS problem
is already obtained variationally and leads to the (in principle)
exact density.
The latter can be used to compute the embedding potential
in the effective Hamiltonian and the
expressions of the per-site energy and the double occupation
[Eqs.~(\ref{eq:per_site_ener}) and
(\ref{eq:dblocc_SOET}), respectively]. This is the strategy employed
in the main text.

\begin{figure}
\centering
\resizebox{\columnwidth}{!}{
\includegraphics[scale=1]{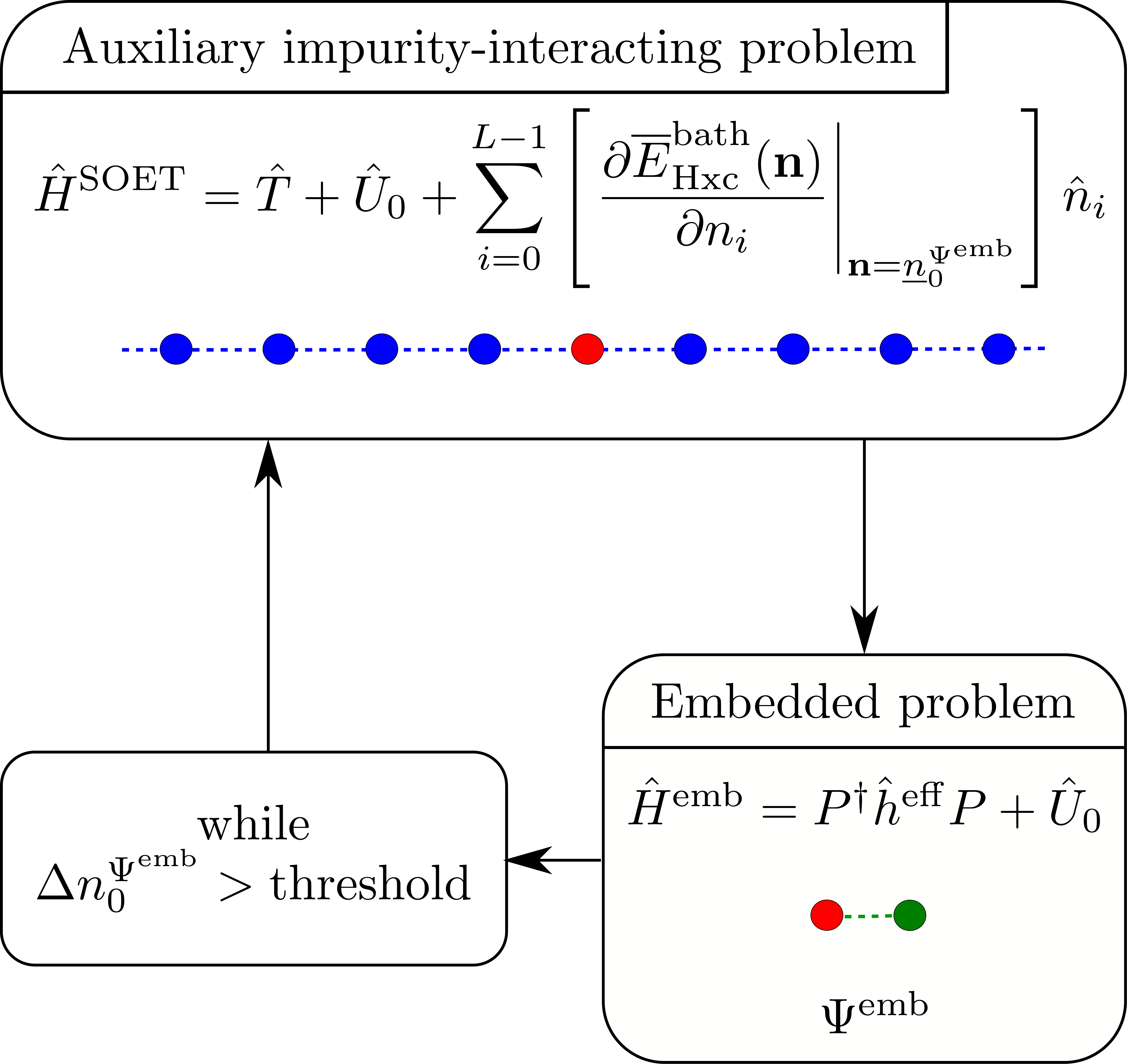}
}
\caption{Self-consistent loop in P-SOET for a single impurity site. The projector is given by the Schmidt decomposition of the KS wavefunction (see Fig.~\ref{fig:PSOET}).}
\label{fig:PSOET_selfcons}
\end{figure}

In practice, approximate 
functionals lead to an approximate
embedding potential that does not guarantee 
the recovering of the exact impurity
occupations. This can be taken into account by
updating the embedding potential
self-consistently with the impurity occupations
of the embedded problem [Eq.~(\ref{eq:occ_emb})].
Note that,
for a uniform model, the knowledge of a single site occupation
is in principle sufficient to determine the embedding potential
exactly, providing that the exact bath Hxc functional
(which itself depends on all sites 
occupation~\cite{senjean2017local}) is known.
One possible and practical
way to reinsert the impurity occupations
into the one-body effective Hamiltonian is to write:
\begin{eqnarray}\label{eq:h_eff_nimp}
\hat{h}^{\rm eff}  = \hat{T} +
\sum_{i=0}^{L-1} \left[\left.
 \dfrac{\partial \overline{E}^{\rm bath}_{{\rm Hxc},M}
 \left(\bfn\right)}{\partial n_i} 
 \right|_{\bfn = \left\lbrace \bfn^{\Psi^{\rm emb}_M}, \bfn^{\rm bath}_M\right\rbrace} \right]
 \hat{n}_i, \nonumber \\
\end{eqnarray}
where
\begin{eqnarray}\label{eq:nimp}
\bfn^{\Psi^{\rm emb}_M} = \left\lbrace n_{0}^{\Psi_M^{\rm emb}}, n_{1}^{\Psi_M^{\rm emb}},
\hdots, n_{M-1}^{\Psi_M^{\rm emb}} \right\rbrace
\end{eqnarray}
is the impurity density (vector of size $M$) and
\begin{eqnarray}\label{eq:nbath}
\bfn^{\rm bath}_M = \left\lbrace \dfrac{1}{M} 
\sum_{i=0}^{M-1} n_{i}^{\Psi_M^{\rm emb}}  \right\rbrace
\end{eqnarray}
is the (uniform)
bath density (vector of size $L - M$), determined by
the mean-average of the impurity occupations.
Note that other
artificial choices (not considered here)
could be made, like keeping the bath occupations frozen
and equal to
the ones obtained from the KS-SOFT calculation.
The embedding potential obtained from the new density
defines a new embedded problem, which we solve 
to obtain a new impurity density. This procedure is iterated until 
convergence of the impurity density is reached, 
as described in Fig.~\ref{fig:PSOET_selfcons}. This self-consistent loop can easily be added to Fig.~\ref{fig:PSOET} for completeness.

One drawback of this self-consistent procedure
is
that the sum of the occupations in $\bfn^{\Psi^{\rm emb}_M}$
and $\bfn^{\rm bath}_M$ are not constrained to sum up to the
total number of electrons.
This could be fixed
by optimizing the embedding potential of
the embedded problem to reproduce the exact
uniform density (obtained by solving the KS problem), 
in the line of DMET or DET for the correlation potential.
For non-uniform models,
one should divide the full system
into multiple fragments, and
the density of the full system would be
rebuilt exactly by combining all the 
fragment occupations.

\subsection{Self-consistent impurity occupations}

\begin{figure}
\centering
\resizebox{\columnwidth}{!}{
\includegraphics[scale=1]{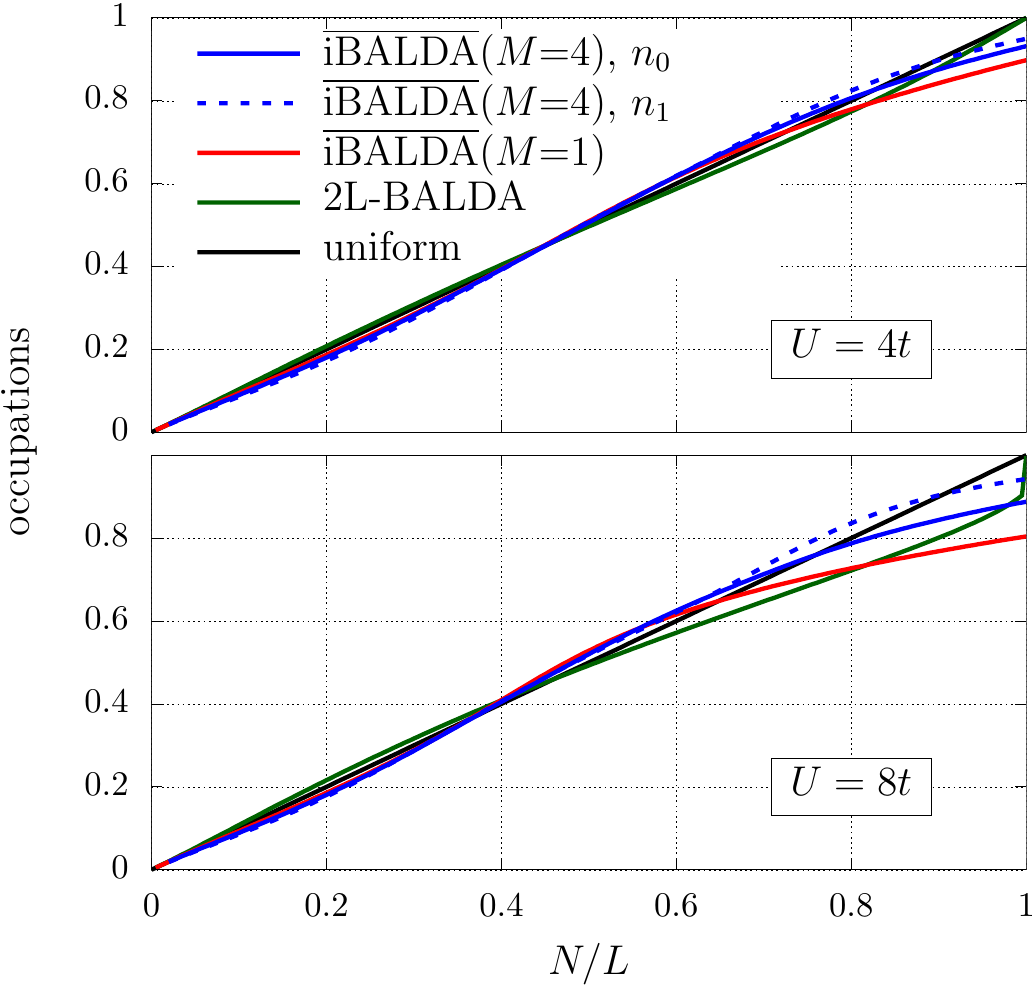}
}
\caption{Impurity occupation(s) obtained 
self-consistently in P-SOET
for different approximations, as a function of the exact uniform
occupation $n = N/L$. We consider a system of
$L = 400$ sites with $U=  4t$ (top panel)
and $U = 8t$ (bottom panel).}
\label{fig:occ}
\end{figure}

The impurity occupations, determined self-consistently
as described in the previous section, are shown in Fig.~\ref{fig:occ} 
for $U = 4t$ (top panel)
and $U=8t$ (bottom panel). Due to the boundary conditions,
the impurity occupations
in $\overline{\rm iBALDA}$($M$=4) are two-by-two equivalent,
e.g. the impurities at the extremities of 
the fragment ($n_0 = n_3$) and
the ones in the middle of the fragment
surrounded 
by the two other impurities ($n_1 = n_2$). 
Therefore, only $n_0$ and $n_1$ are represented.
As readily seen in Fig.~\ref{fig:occ},
the deviation from the exact density
is significant in the vicinity of half-filling. This would lead to
strong density-driven errors in P-SOET in this regime. 
Note however that the 2L-BALDA occupation
for $U=4t$ is accurate in the entire density domain.
The deviations then increase for $U = 8t$, showing that it is
harder to converge to the correct density as the correlation strength
increases.

\begin{figure}
\centering
\resizebox{\columnwidth}{!}{
\includegraphics[scale=1]{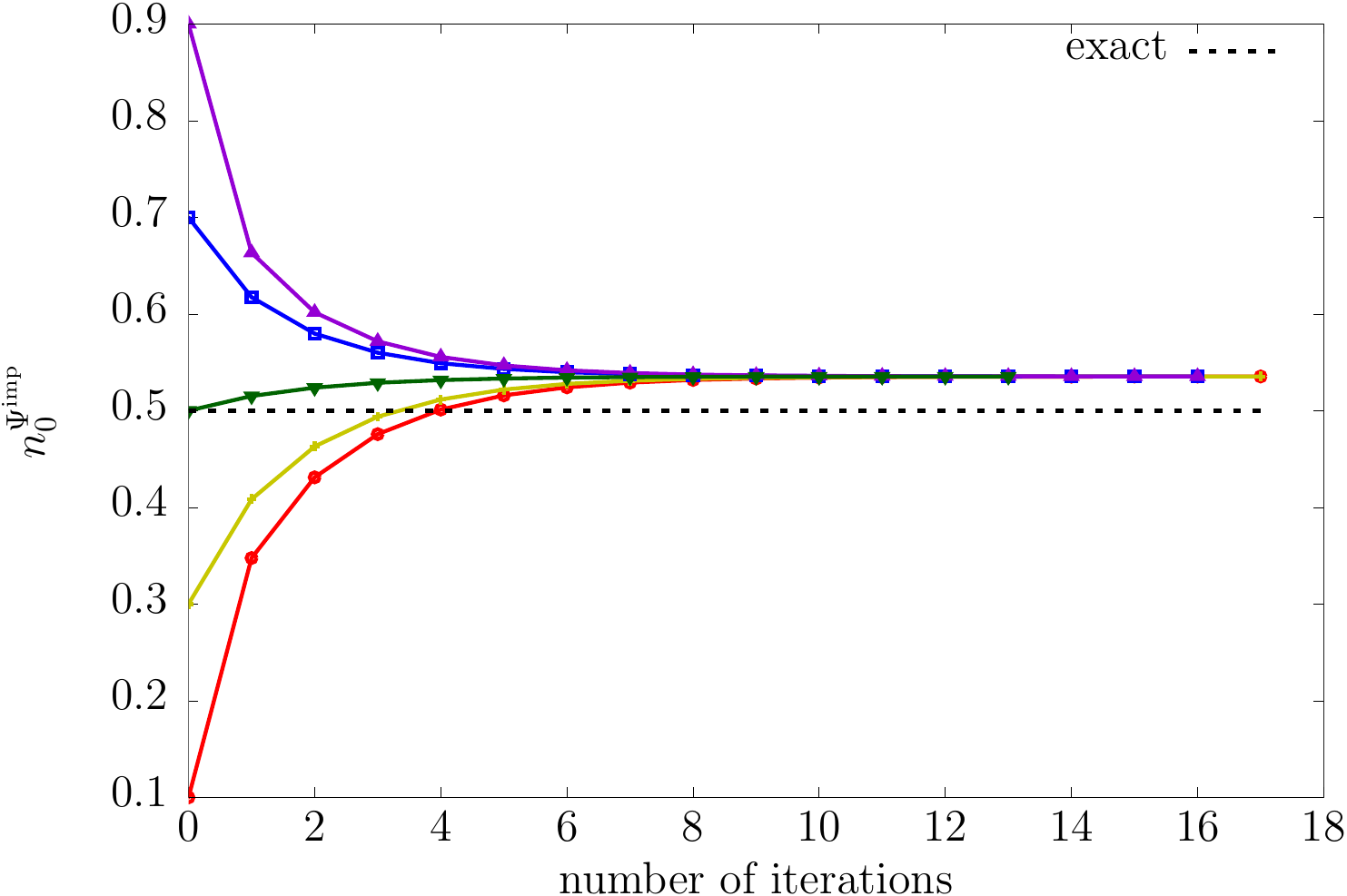}
}
\caption{Impurity occupation
within $\overline{\rm iBALDA}$($M$=1) at each iteration
of the self-consistent procedure, starting from different uniform
densities. The $L = 400$ sites model is considered,
with $N = 200$ electrons and 
$U = 10t$.}
\label{fig:guess}
\end{figure}

Interestingly, the occupations in Fig.~\ref{fig:occ}
are
really similar to the ones obtained from
SOET (Figs.~7 and 8 of Ref.~\cite{senjean2018multiple}). 
This demonstrates the stability of this self-consistent procedure,
which was not obvious considering the results of 
Table~\ref{tab:LFpot} showing that the impurity occupation
obtained in P-SOET was not guaranteed to be the same as in SOET.
Therefore, it seems that the self-consistent P-SOET
is stable and gives similar results than SOET. 
To check this stability further, we start with 
other initial densities than the
one obtained from the KS-SOFT calculation
to determine the embedding potential
in the effective problem [Eq.~(\ref{eq:h_eff})].
Fig.~\ref{fig:guess} shows the impurity occupation
within $\overline{\rm iBALDA}$($M$=1)
at each iteration of the self-consistent procedure,
starting from different initial densities. It is
clear
that the impurity occupation converges to the same value
irrespective of the initial settings. Also, only
a few number of iterations are needed, just like in
self-consistent DMET but without the expensive fitting of the
correlation potential at each iteration (for
which alternatives have been recently
developed~\cite{wu2019projected}).
The difference between the converged occupation and the exact
uniform one in Fig.~\ref{fig:guess} 
is mainly due to the use of approximate
functionals, which determine the
embedding potential
[Eq.~(\ref{eq:embedding_pot})].

It is worth mentioning that convergence issues arise for 
$U \leqslant 1.8t$ at half-filling.
Indeed, when the occupation is too close to
1, fluctuations of the occupations around 1 are known to appear
due to the discontinuity in the
potential~\cite{lima2003density,senjean2018multiple}, as discussed
in Sec.~\ref{sec:approx}.
For $U \geqslant 1.8t$ the impurity occupations converge sufficiently 
far away from 1 to avoid this issue.
Appart from that, we have not observed any convergence problem in
this self-consistent formulation of P-SOET.

\end{document}